\documentclass[12pt,draftcls,peerreview,onecolumn]{IEEEtran}

\ifCLASSINFOpdf
  \usepackage[pdftex]{graphicx}
\else
\fi
\usepackage[cmex10]{amsmath}
\usepackage[tight,footnotesize]{subfigure}

\usepackage{amsfonts,amsmath,amsthm,amssymb,graphicx,epstopdf,cite,subfigure}
\usepackage{psfrag,pifont,cite,graphics,epsfig,amscd,helvet,multirow,enumerate}
\usepackage{color}
\usepackage{epstopdf}
\usepackage{accents}
\usepackage{psfrag,threeparttable,url}
\usepackage{mathtools}
\usepackage{multirow}
\usepackage[linesnumbered,ruled,norelsize]{algorithm2e}
	\makeatletter
	\newcommand{\removelatexerror}{\let\@latex@error\@gobble}
	\makeatother

\usepackage{booktabs}
\usepackage{pgfplots}
\usepackage{epstopdf}

\newtheorem{theorem}{Theorem}[section]
\newtheorem{lemma}[theorem]{Lemma}

\newtheorem{definition}[theorem]{Definition}

\newtheorem{remark}[theorem]{Remark}
\begin{document}

\newcommand{\matc}[2][ccccccccccccccccccc]{\left[
\begin{array}{#1}
#2\\
\end{array}
\right]}
\newcommand{\matr}[2][rrrrrrrrrrrrrrrrrrrrrrrr]{\left[
\begin{array}{#1}
#2\\
\end{array}
\right]}
\newcommand{\matl}[2][lllllllllllllllllll]{\left[
\begin{array}{#1}
#2\\
\end{array}
\right]}

\newcommand{\pt}[2]{P_{T_{\mathbf{Y}_{#1}}\widetilde{\mathcal{Y}}}(#2)}

\newcommand*\circled[1]{\tikz[baseline=(char.base)]{
  \node[shape=circle,draw,inner sep=2pt] (char) {#1};}}

%%%%%%%%%%%%%%%%%%%%%%%%%%%%%%%%%%%%%%%%%%%%%%%%%%%%%%%%%%%%%

{\Large \textbf{Notice:} This work has been submitted to the IEEE for possible publication. Copyright may be transferred without notice, after which this version may no longer be accessible.}

\clearpage

%%%%%%%%%%%%%%%%%%%%%%%%%%%%%%%%%%%%%%%%%%%%%%%%%%%%%%%%%%%%%

\title{Localization of IoT Networks Via Low-Rank Matrix Completion}

%% author names and affiliations
%% use a multiple column layout for up to three different
%% affiliations
%\author{\IEEEauthorblockN{Luong T. Nguyen, Junhan Kim, Sangtae Kim, Byonghyo Shim}
%
%\IEEEauthorblockA{Information System Laboratory\\
%Department of Electrical and Computer Engineering, Seoul National University\\
%Email: \{ltnguyen, junhankim, stkim, bshim\}@islab.snu.ac.kr}
%\thanks{This work was sponsored by the National Research Foundation of Korea (NRF) grant funded by the Korean government (MSIP) (2016R1A2B3015576) and Electronics and Telecommunications Research Institute (ETRI) grant funded by the Korean government [16ZI1100, Wireless Transmission Technology in Multi-point to Multi-point Communications].}
%\thanks{A part of this paper was presented at the Information Theory and Applications (ITA) workshop \cite{localization:luongITA} and International Conference on Communications in China (ICCC), 2016 \cite{localization:luongICCC}.}
%}

\author{Luong T. Nguyen,
	Junhan Kim,
	Sangtae Kim,
	and~Byonghyo Shim,~\IEEEmembership{Senior Member,~IEEE}% <-this % stops a space
	%\thanks{The Department
	%of Electrical and Computer Engineering, Seoul National University, Korea.}% <-this % stops a space
	\thanks{L. T. Nguyen, J. Kim, S. Kim, and B. Shim are with the Department of Electrical and Computer Engineering, Seoul National University, South Korea (Email: \{ltnguyen, junhankim, stkim, bshim\}@islab.snu.ac.kr).
	
		A part of this paper was presented at the Information Theory and Applications (ITA) workshop \cite{localization:luongITA} and International Conference on Communications in China (ICCC), 2016 \cite{localization:luongICCC}.
		
		%	This work was supported by `The Cross-Ministry Giga KOREA Project' grant funded by the Korea government (MSIT) (No.GK17P0501, Development of Ultra Low-Latency Radio Access Technologies for 5G URLLC Service) and the National Research Foundation of Korea(NRF) grant funded by the Korea government(MSIT) (No. NRF-2016K1A3A20006019). 
		This work was supported by the National Research Foundation of Korea (NRF) grant funded by the Korean government (MSIP) (2014R1A5A1011478) and the MSIT (Ministry of Science and ICT), Korea, under the ITRC (Information Technology Research Center) support program (IITP-2019-2017-0-01637) supervised by the IITP (Institute for Information \& Communications Technology Promotion).

	}
}


%Telephone: (800) 555--1212, Fax: (888) 555--1212}

\markboth{IEEE TRANSACTIONS ON COMMUNICATIONS,~Vol.~X, No.~Y, May~2019}%
{Shell \MakeLowercase{\textit{et al.}}: Bare Demo of IEEEtran.cls for IEEE Communications Society Journals}


\maketitle
%\IEEEpeerreviewmaketitle

\begin{abstract}
Location awareness, providing ability to identify the location of sensor, machine, vehicle, and wearable device, is a rapidly growing trend of hyper-connected society and one of key ingredients for internet of things (IoT) era. In order to make a proper reaction to the collected information from {\it things}, location information of things should be available at the data center. One challenge for the IoT networks is to identify the location map of whole nodes from partially observed distance information. An aim of this paper is to present an algorithm to recover the Euclidean distance matrix (and eventually the location map) from partially observed distance information. By casting the low-rank matrix completion problem into the unconstrained minimization problem in a Riemannian manifold in which a notion of differentiability can be defined, we solve the low-rank matrix completion problem using a modified conjugate gradient algorithm.  
From the convergence analysis, we show that LRM-CG converges linearly to the original Euclidean distance matrix under the extended Wolfe's conditions. From the numerical experiments, we demonstrate that the proposed method, called {\it localization in Riemannian manifold using conjugate gradient} (LRM-CG), is effective in recovering the Euclidean distance matrix. 
\end{abstract}


% \fi
\IEEEpeerreviewmaketitle

\section{Introduction}
\label{sec:sec1}
Recently, Internet of Things (IoT) has received much attention for its plethora of applications, such as healthcare, surveillance, automatic metering, and environmental monitoring. In sensing the environmental data (e.g. temperature, humidity, pressure, pollution density, and object movements), wireless sensor network consisting of thousands of more sensor nodes is popularly used \cite{localization:delamo2015,localization:sunwoo2014,localization:hackmann2014,localization:pal2010}. In order to make a proper reaction to the collected environmental data, location information of sensor nodes should be available at the data center (basestation) \cite{localization:hodge2014,localization:rawat2014}. Since actions in IoT networks, such as fire alarm, energy transfer, and emergency request, are made primarily on the data center, an approach to identify the location information of whole nodes at the data center has received much attention. In this so-called \textit{network localization} (a.k.a. cooperative localization), each node measures the distance information of adjacent nodes and then forwards it to the data center \cite{localization:aspnes2006}. Then the data center constructs a map of sensor nodes using the collected distance information \cite{localization:shang2003}. In obtaining the distance information, various measuring modalities, such as received signal strength indication (RSSI) \cite{localization:parker2007}, time of arrival (ToA) \cite{localization:dardari2008}, time difference of arrival (TDoA) \cite{localization:zhang2005}, and angle of arrival (AoA) \cite{localization:pal2010}, have been employed. These approaches are simple and effective in measuring the short-range distance and also commonly used for indoor environments.

When it comes to the network localization in IoT, there are two major tasks to be done. First, distance information should be converted to the location information. Typically, converted location information of the sensor node is local, meaning that the location information is true in the relative sense. Thus, proper adjustment of the location information is needed to obtain the absolute (true) location information. In fact, since the local location of a sensor node might be different from the absolute location by some combinations of translations, rotations, and reflections, absolute locations of a few sensor nodes (\textit{anchor nodes}) are needed to transform the local locations into the absolute locations. It has been shown that when the number of anchor nodes is enough (e.g., four anchor nodes in $\mathbb{R}^{2}$), one can easily identify the absolute locations of sensor nodes \cite{localization:shang2003}.  
Readers are referred to  \cite{localization:hodge2014,localization:rawat2014} for more details.

%----------------------------------------------------------
\begin{figure*} [t] 
	\centering
  {\epsfig{figure=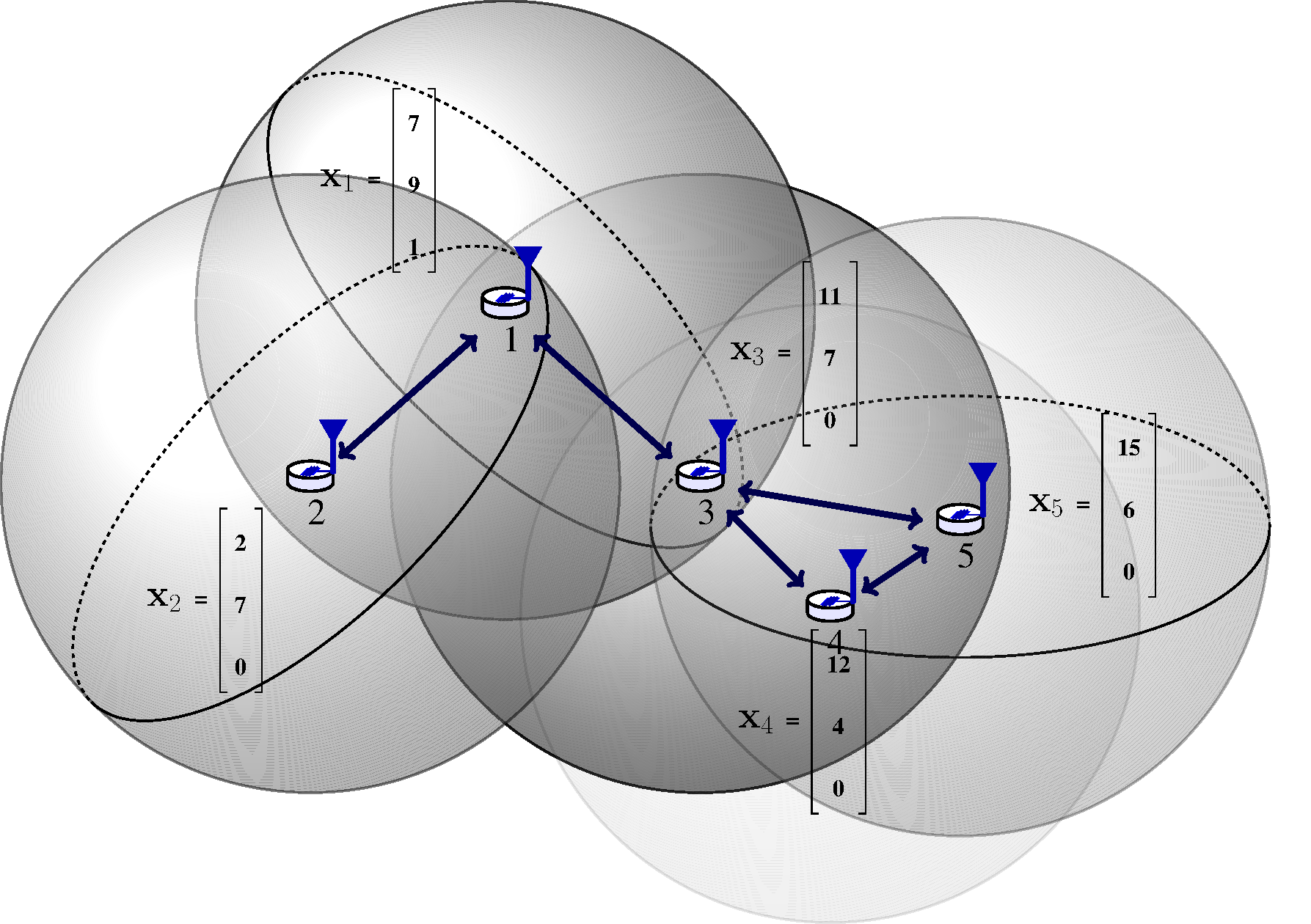, scale=0.45}}
  \caption {Sensor nodes deployed to measure not only environment information but also their pairwise distances. The observed distances are represented by two-sided arrows. The shadow spheres represent the radio communication range of the sensor nodes.}  
\label{fig:fig1}
\end{figure*}
%----------------------------------------------------------

Second and more important problem is that the data center does not have enough distance information to identify the locations of sensor nodes. For various reasons, such as the power outage of a sensor node or the limitation of radio communication range, only small number of distance information is available at the data center. This situation can also occur in the hierarchical or relay-based IoT networks where only intermediate or cluster head node sends the distance information to the data center. Also, in the vehicular networks it might not be possible to measure the distance of all adjacent vehicles when a vehicle is located at the dead zone. Similar behavior can also be observed in underwater acoustic communication environments. To illustrate this scenario, we depict a simple network consisting of five sensor nodes in Fig. \ref{fig:fig1}. We see that only a small number of pairwise distances are measured, and hence there are many unknown entries in the observation matrix $\mathbf{D}_{o}$: 
\begin{equation}
\mathbf{D}_{o} = \matc{0 & d_{12}^2 & d_{13}^2 & ?  & ?\\ 
      d_{21}^2 & 0 & ? & ? & ?\\
      d_{31}^2 & ? & 0 & d_{34}^2 & d_{35}^2\\
      ? & ? & d_{43}^2 & 0 & d_{45}^2\\
      ? & ? & d_{53}^2 & d_{54}^2 & 0}\nonumber,
\end{equation}
where the question mark $?$ indicates unknown entries of $\mathbf{D}$. 
In finding out the node location using the observed matrix $\mathbf{D}_o$, multidimensional scaling (MDS) technique has been employed \cite{localization:shang2003}. In a nutshell, MDS reconstructs the Euclidean distance matrix using the shortest path algorithm and then computes the node locations using a truncated eigendecomposition. Semidefinite programming (SDP) technique using a convex relaxation of nonconvex quadratic constraints of the node locations has also been used to this \cite{localization:guo2016, localization:biswas2004}. However, SDP-based techniques are computationally expensive since the computational complexity depends heavily on the problem size \cite{localization:biswas2004}.

As an alternative approach, matrix completion techniques reconstructing the Euclidean distance matrix $\mathbf{D}$ using partially observed entries have been proposed in recent years\cite{localization:candes2009,
localization:cai2010, localization:lin2010, localization:mishra2011, localization:jain2013}. 
In general, one cannot recover the original matrix from a knowledge of a subset of its entries since there are infinitely many completion options for the unknown entries. However, if a matrix is low-rank, then it can be recovered from the partial observation matrix \cite{localization:candes2009}. Since the rank of the Euclidean distance matrix $\mathbf{D}$ in the $k$-dimensional Euclidean space is at most $k+2$ ($k = 2$ or $3$) \cite{localization:dattorro2005}, it can be readily modeled as a low-rank matrix. The problem to recover a low-rank matrix $\mathbf{D}$ from the small number of known entries can be expressed as 
\begin{equation}
\begin{matrix}
\min\limits_{\widetilde{\mathbf{D}}\:\in\:\mathbb{R}^{n\times n}} & \frac{1}{2}\|\mathcal{P}_E(\widetilde{\mathbf{D}})-\mathcal{P}_E(\mathbf{D}_{o})\|_F^2,\\
\text{s.t.} & \text{rank}(\widetilde{\mathbf{D}}) \leq k+2,
\end{matrix}  
\label{eq:eq209} 
\end{equation}
where $\mathcal{P}_E$ is the sampling operator given by 
\begin{equation}
[\mathcal{P}_E(\mathbf{A})]_{ij} = \left\lbrace\begin{matrix}
A_{ij} & \text{if } (i,j)\in E\\
0 & \text{otherwise}.
\end{matrix} \right. \nonumber
\end{equation}
In the RSSI-based distance model, for example, $E = \{(i,j):\|\mathbf{x}_i-\mathbf{x}_j\|_2\leq r\}$ would be the set of observed indices for a given radio communication range $r$. 
Note that this problem is robust to the observation error and noise since it uses the Frobenius norm-based cost function. 
Also, this approach is good fit for the situation where the rank constraint is known in a priori, which is true in our case. In recent years, various approaches to find a solution of (\ref{eq:eq209}) have been suggested. In \cite{localization:candes2009}, the nuclear norm minimization (NNM) has been proposed. In \cite{localization:cai2010}, a singular value thresholding (SVT) technique that shrinks the number of singular values and as a result ensures the low rank structure of the output matrix has been proposed. As extensions of SVT technique, the augmented Lagrange multiplier (ALM) and accelerated proximal gradient (APG) algorithms have also been proposed in \cite{localization:lin2010} and \cite{localization:lin2009}.

An aim of this paper is to propose a Euclidean distance matrix completion technique for the IoT network localization. In our approach, we express the Euclidean distance matrix $\mathbf{D}$ as a function of the low rank positive semidefinite (PSD) matrix. Since the set of these matrices forms a Riemannian manifold in which the notation of differentiability can be defined, we can recycle, after a proper modification, an algorithm in the Euclidean space. In order to solve the problem (\ref{eq:eq209}), we propose a modified conjugate gradient algorithm, referred to as {\it localization in Riemannian manifold using conjugate gradient} (LRM-CG).
The main contributions of this paper are as follows:
\begin{itemize}
\item We propose a matrix completion-based IoT network localization algorithm called LRM-CG. Our numerical and simulation results demonstrate that LRM-CG can exactly recover the Euclidean distance matrix from partial measurements, achieving MSE $\leq 10^{-5}$ using 40\% of measurements (see Subsection \ref{subsec:subsec3}). 
\item We propose an extension of LRM-CG to cope with the scenario in which the observed pairwise distances are contaminated by the outliers. By modeling outliers as a sparse matrix and then adding a regularization term of the outlier matrix into the Frobenius norm-based problem, we can effectively control the outliers. From simulation results, we observe that the extended LRM-CG is effective in handling the outliers, achieving mean square localization error being less than 0.5m up to 20\% outlier ratio (see Subsection \ref{subsec:subsec4}).        
\end{itemize}
        
We briefly summarize notations used in this paper. $<\boldsymbol\beta_1,\boldsymbol\beta_2>$ is the inner product between $\boldsymbol\beta_1$ and $\boldsymbol\beta_2$, i.e., $<\boldsymbol\beta_1,\boldsymbol\beta_2> = \text{tr}(\boldsymbol\beta_1^T\boldsymbol\beta_2)$. $\text{diag}(\mathbf{A})$ is the vector formed by the main diagonal of a matrix $\mathbf{A}$. $\text{Sym}(\mathbf{A})$ and $\text{Skew}(\mathbf{A})$ are the matrices formed by $\text{Sym}(\mathbf{A}) = \frac{1}{2}(\mathbf{A}+\mathbf{A}^T)$ and $\text{Skew}(\mathbf{A}) = \frac{1}{2}(\mathbf{A}-\mathbf{A}^T)$ for any square matrix $\mathbf{A}$, respectively. Note that $\mathbf{A} = \text{Sym}(\mathbf{A})+\text{Skew}(\mathbf{A})$. $\text{eye}(\mathbf{a})$ is the diagonal matrix formed by $\mathbf{a}$. For an orthogonal matrix $\mathbf{Q}\in\mathbb{R}^{n\times k}$ with $n\textgreater k$, we define its orthogonal complement $\mathbf{Q}_\perp\in\mathbb{R}^{n\times (n-k)}$ such that $\matc{\mathbf{Q} & \mathbf{Q}_\perp}$ forms an orthonormal matrix. Given a function $f:\mathbf{Y}\in\mathbb{R}^{n\times n}\rightarrow f(\mathbf{Y})\in\mathbb{R}$, $\nabla_{\mathbf{Y}}f(\mathbf{Y})$ is the Euclidean gradient of $f(\mathbf{Y})$ with respect to $\mathbf{Y}$, i.e., $\left[\nabla_{\mathbf{Y}}f(\mathbf{Y})\right]_{ij}=\frac{\partial f(\mathbf{Y})}{\partial y_{ij}}$. For a given matrix $\mathbf{A}=\matc{\mathbf{a}_1 & \mathbf{a}_2 & \cdots & \mathbf{a}_n}\in\mathbb{R}^{n\times n}$, the vectorization of $\mathbf{A}$, denoted by $\text{vec}(\mathbf{A})$, is defined as $\text{vec}(\mathbf{A}) = \matc{\mathbf{a}_1^T & \mathbf{a}_2^T & \cdots & \mathbf{a}_n^T}^T$. $\{\mathbf{e}_{i}\}_{i=1}^n$ are the $n\times 1$ standard basis vectors of $\mathbb{R}^n$. $\mathbf{A}\odot\mathbf{B}$ is the Hadamard product of two matrices $\mathbf{A}$ and $\mathbf{B}$. $\mathbf{1} = \matc{1 & 1 & \cdots & 1}^T$ is all-ones vector.

\section{The LRM-CG Algorithm}
\label{sec:sec2} 

In this section, we present the proposed LRM-CG algorithm. By exploiting the smooth Riemannian manifold structure of the set of the low-rank symmetric PSD matrices, we formulate the matrix completion problem (\ref{eq:eq209}) as an unconstrained optimization problem on the smooth Riemannian manifold. Roughly speaking, smooth manifold is a generalization of the Euclidean space on which a notion of differentiability exists. For more rigorous definition, see, e.g.,  \cite{localization:absil2008}. A smooth manifold together with an inner product, often called a Riemannian metric, forms a smooth Riemannian manifold. Since the smooth Riemannian manifold is a differentiable structure equipped with an inner product, we can use various ingredients such as Riemannian gradient, Hessian matrix, exponential map, and parallel translation, for solving optimization problems with quadratic cost function \cite{localization:absil2008}. Therefore, optimization techniques in the Euclidean vector space (e.g., steepest descent, Newton method, conjugate gradient method) can be readily extended to solve a problem in the smooth Riemannian manifold.

\subsection{Problem Model}
We consider the problem of $n$ sensor nodes distributed in the $k$-dimensional Euclidean space. Let $\mathbf{x}_i$ be the coordinate vector of the $i$-th sensor $(1\leq i \leq n)$ and $\mathbf{X} = [ \mathbf{x}_{1} \ \mathbf{x}_{2} \ \cdots \ \mathbf{x}_{n}]^T$. Then, the distance $d_{ij}$ between the $i$-th and $j$-th sensors is given by $d_{ij}^{2} = \| \mathbf{x}_{i} - \mathbf{x}_{j} \|_{2}^{2} = \mathbf{x}_i^T\mathbf{x}_i + \mathbf{x}_j^T\mathbf{x}_j - 2\mathbf{x}_i^T\mathbf{x}_j$, and thus the Euclidean distance matrix $\mathbf{D}$ satisfies 
\begin{equation}
\mathbf{D} = g(\mathbf{XX}^T),
\label{eq:eq211}
\end{equation} 
where $g(\mathbf{XX}^T) = 2\text{Sym}(\text{diag}(\mathbf{XX}^T)\mathbf{1}^T - \mathbf{XX}^T)$. In the example illustrated in Fig. \ref{fig:fig1}, we have 
\begin{align*}
\mathbf{X} & = \matc{\mathbf{x}_1 & \mathbf{x}_2 & \mathbf{x}_3 & \mathbf{x}_4 & \mathbf{x}_5}^T \notag\\
& = \matc{7 & 2 & 11 & 12 & 15 \\ 9 & 7 & 7 & 4 & 6 \\ 1 & 0 & 0 & 0 & 0}^T, 
\end{align*}
and
\begin{equation}
\mathbf{D} = g(\mathbf{XX}^T) = \matc{0  &  30  &  21  &  51  &  74\\
    30  &   0  &  81 &  109 &  170\\
    21  &  81  &   0  &  10  &  17\\
    51 &  109  &  10  &   0  &  13\\
    74  & 170  &  17  &  13  &   0}.\nonumber
\end{equation}
The problem to identify the node locations from a partial observation of $\mathbf{D}$ is formulated as
\begin{equation}
\begin{matrix}
\min\limits_{\widetilde{\mathbf{X}}\:\in\:\mathbb{R}^{n\times k}} & \frac{1}{2}\|\mathcal{P}_E(g(\widetilde{\mathbf{X}}\widetilde{\mathbf{X}}^T))-\mathcal{P}_E(\mathbf{D}_{o})\|_F^2 .
\label{eq:eq502}
\end{matrix}  
\end{equation}
When $n$ nodes ($n\geq k$) are distributed in $k$-dimensional Euclidean space, $\text{rank}(\mathbf{D})\leq k+2$ \cite{localization:guo2016,localization:dattorro2005}. 
Incorporating this constraint, we have
\begin{equation}
\begin{matrix}
\min\limits_{\widetilde{\mathbf{D}}\:\in\:\mathbb{R}^{n\times n}} & \frac{1}{2}\|\mathcal{P}_E(\widetilde{\mathbf{D}})-\mathcal{P}_E(\mathbf{D}_{o})\|_F^2,\\
\text{s.t.} & \text{rank}(\widetilde{\mathbf{D}}) \leq k+2.
\end{matrix}  
\label{eq:eq763}
\end{equation}  
Here, we use $\widetilde{\mathbf{X}}$ and $\widetilde{\mathbf{D}}$ as the optimization variables of the problems~\eqref{eq:eq502} and~\eqref{eq:eq763}, respectively, to differentiate them from the true coordinate matrix $\mathbf{X}$ and the true distance matrix $\mathbf{D}$.
In order to suppress the effect of large magnitude errors, we can incorporate a weight matrix $\mathbf{W}$ into (\ref{eq:eq763})\footnote{If the observed entries are accurate, we simply set $w_{ij} = 1$ for all $(i,j)\in E$. However, in many practical scenarios where range-based techniques are employed, the measurement accuracy might be inversely proportional to the magnitude of the observed distances \cite{localization:jianwu2009}, which needs to be accounted for the choice of $w_{ij}$.}. Thus,
\begin{equation}
\begin{matrix}
\min\limits_{\widetilde{\mathbf{D}}\:\in\:\mathbb{R}^{n\times n}} & \frac{1}{2}\|\mathbf{W}\odot(\mathcal{P}_E(\widetilde{\mathbf{D}})-\mathcal{P}_E(\mathbf{D}_{o}))\|_F^2,\\
\text{s.t.} & \text{rank}(\widetilde{\mathbf{D}}) \leq k+2,
\end{matrix}  
\label{eq:eq511}
\end{equation}
where $w_{ij}$ is the $(i,j)$-th entry of $\mathbf{W}$ satisfying $w_{ij}\textgreater 0$ for $(i,j)\in E$ and zero otherwise. 
Noting that $\widetilde{\mathbf{D}} = g(\mathbf{Y})$ for a PSD matrix $\mathbf{Y}$, we further have
\begin{equation}
\begin{matrix}
\min\limits_{\mathbf{Y}\:\in\:\mathcal{Y}} & \frac{1}{2}\|\mathbf{W}\odot(\mathcal{P}_E(g(\mathbf{Y}))-\mathcal{P}_E(\mathbf{D}_{o}))\|_F^2 ,
\end{matrix}  
\label{eq:eq503}
\end{equation}
where $\mathcal{Y} = \{\widetilde{\mathbf{X}}\widetilde{\mathbf{X}}^T:\widetilde{\mathbf{X}}\in\mathbb{R}^{n\times k}\}$\footnote{Note that the feasible set $\mathcal{Y}$ includes the rank constraint $\text{rank}(\widetilde{\mathbf{D}}) \leq k+2$.}.

%----------------------------------------------------------
\begin{figure*} [t]
 \centering
    \subfigure[]
  {\epsfig{figure=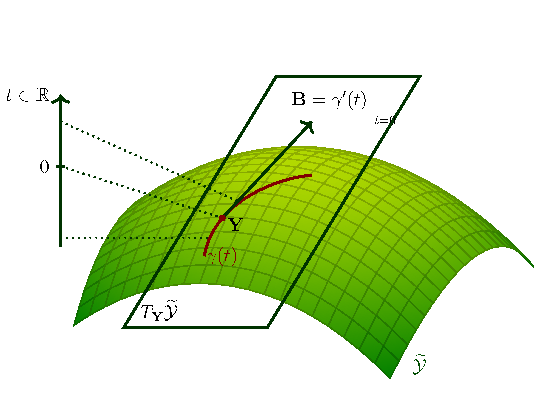}}
	\subfigure[]
  {\epsfig{figure=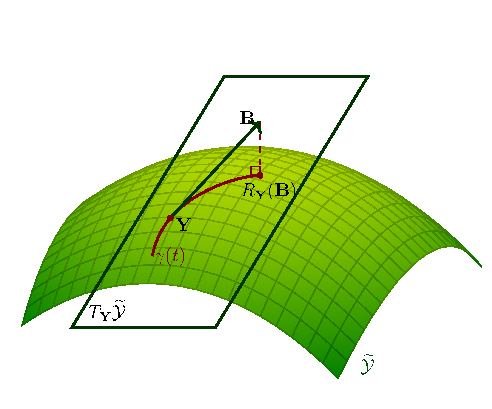}}
  \caption {Illustration of (a) the tangent space $T_{\mathbf{Y}}\widetilde{\mathcal{Y}}$ and (b) the retraction operator $R_{\mathbf{Y}}$ at a point $\mathbf{Y}$ in the embedded manifold $\widetilde{\mathcal{Y}}$.}  
\label{fig:fig2}
\end{figure*}
%----------------------------------------------------------

When the nodes are randomly distributed in $k$-dimensional Euclidean space, rank of the location matrix $\mathbf{X}$ is $k$ almost surely\footnotemark\footnotetext{Consider the case that sensor nodes are randomly distributed in 2D Euclidean space, then $\text{rank}(\mathbf{X}) = 1$ if and only if all of nodes are co-linear. This event happens if there exists a constant $\rho$ such that $x_{i1} = \rho x_{i2}$ for any $i$-th row. The probability of this event $\prod\limits_{i=1}^nP(x_{i1}=\rho x_{i2}) = [P(x_{11}=\rho x_{12})]^n$ is negligible when the number of sensor nodes are sufficiently large.}. Thus, we can strengthen the constraint set from $\mathcal{Y}$ to $\widetilde{\mathcal{Y}} = \{\widetilde{\mathbf{X}}\widetilde{\mathbf{X}}^T:\widetilde{\mathbf{X}}\in\mathbb{R}^{n\times k},\:\text{rank}(\widetilde{\mathbf{X}})=k\}$, and thus
\begin{equation}
\begin{matrix}
\min\limits_{\mathbf{Y}\:\in\:\widetilde{\mathcal{Y}}} & \frac{1}{2}\|\mathbf{W}\odot(\mathcal{P}_E(g(\mathbf{Y}))-\mathcal{P}_E(\mathbf{D}_{o}))\|_F^2 ,
\end{matrix}  
\label{eq:eq520}
\end{equation}
In the sequel, we denote $f(\mathbf{Y}) = \frac{1}{2}\|\mathbf{W}\odot(\mathcal{P}_E(g(\mathbf{Y}))-\mathcal{P}_E(\mathbf{D}_{o}))\|_F^2$ for notational simplicity. Once the solution $\mathbf{Y}^\ast$ of the problem (\ref{eq:eq520}) is obtained, we can recover the node location using the eigen-decomposition of $\mathbf{Y}^\ast$ (see Subsection \ref{subsec:subsec002} for details).

\subsection{Optimization over Riemannian Manifold}
\label{subsec:subsec2} 
Let $\mathcal{S} = \{\mathbf{S}\in\mathbb{R}^{n\times k}:\mathbf{S}^T\mathbf{S}=\mathbf{I}_k\}$\footnotemark\footnotetext{$\mathcal{S}$ is an orthogonal Stiefel manifold embedded in $\mathbb{R}^{n\times k}$ \cite{localization:absil2008}.} and $\mathcal{L} = \{\text{eye}([\begin{matrix}
\lambda_1 & \cdots & \lambda_k
\end{matrix} ]^T):\lambda_1\geq \lambda_2\geq \cdots \geq \lambda_k\textgreater 0\}$. Then, for given $\mathbf{T}\in\widetilde{\mathcal{Y}}$, we can express $\mathbf{T}=\mathbf{S}\boldsymbol\Gamma\mathbf{S}^T$ using the eigenvalue decomposition and thus  
\begin{equation}
\widetilde{\mathcal{Y}} = \{\mathbf{S}\boldsymbol\Gamma\mathbf{S}^T: \mathbf{S}\in \mathcal{S} , \boldsymbol\Gamma\in\mathcal{L}\}, 
\label{eq:eq217}
\end{equation}
where $\widetilde{\mathcal{Y}}$ is a smooth Riemannian manifold \cite[Ch.5]{localization:helmke1994}. Our approach to solve the problem in a smooth Riemannian manifold is beneficial in two major respects: First, one can easily compute the gradient of the cost function in (\ref{eq:eq520}) using the matrix calculus. Second, one can extend techniques in the Euclidean space to solve the problem (\ref{eq:eq520}).

Since our work relies to a large extent on properties and operators of differential geometry, we briefly introduce tools and ingredients to describe the proposed algorithm. Since $\widetilde{\mathcal{Y}}$ is an embedded manifold in the Euclidean space $\mathbb{R}^{n\times n}$, its tangent spaces are determined by the derivative of its curves, where the curve $\gamma$ of $\widetilde{\mathcal{Y}}$ is a mapping from $\mathbb{R}$ to $\widetilde{\mathcal{Y}}$. 
%Note that the tangent space on the smooth Riemannian manifold is a generalization of the notion of tangent hyperplane in Euclidean space. 
Put it formally, for a given point $\mathbf{Y} = \mathbf{Q}\boldsymbol\Lambda\mathbf{Q}^T\in\widetilde{\mathcal{Y}}$, the tangent space of $\widetilde{\mathcal{Y}}$ at $\mathbf{Y}$, denoted $T_\mathbf{Y}\widetilde{\mathcal{Y}}$, is defined as $T_\mathbf{Y}\widetilde{\mathcal{Y}} = \{\gamma^\prime(0):\gamma \text{ is a curve in }\widetilde{\mathcal{Y}}, \gamma(0) = \mathbf{Y}\}$ (see Fig. \ref{fig:fig2}). The tangent space $T_{\mathbf{Y}}\widetilde{\mathcal{Y}}$ can be expressed as \cite{localization:bart2013}
\begin{eqnarray}
T_{\mathbf{Y}}\widetilde{\mathcal{Y}} & = & \left\lbrace\left[\begin{matrix}
\mathbf{Q} & \mathbf{Q}_{\perp}
\end{matrix} \right]\left[\begin{matrix}
\mathbf{C}_1 & \mathbf{C}_2^T\\
\mathbf{C}_2 & \mathbf{0}
\end{matrix} \right]\left[\begin{matrix}
\mathbf{Q}^T\\
\mathbf{Q}_{\perp}^T
\end{matrix} \right]:\right.\nonumber\\
& & \left. \mathbf{C}_1^T = \mathbf{C}_1\in\mathbb{R}^{k\times k}, \mathbf{C}_2\in\mathbb{R}^{(n-k)\times k}     \right\rbrace . 
\label{eq:eq512}
\end{eqnarray}

A metric on the tangent space $T_{\mathbf{Y}}\widetilde{\mathcal{Y}}$ is defined as the matrix inner product $<\mathbf{B}_1,\mathbf{B}_2 > = \text{tr}(\mathbf{B}_1^T\mathbf{B}_2)$ between two tangent components $\mathbf{B}_1,\mathbf{B}_2\in T_{\mathbf{Y}}\widetilde{\mathcal{Y}}$. Next, we define the orthogonal projection of a matrix $\mathbf{A}$ onto the tangent space $T_{\mathbf{Y}}\widetilde{\mathcal{Y}}$, which will be used to find the closed-form expression of Riemannian gradient in Subsection \ref{subsec:subsec002}. 
\begin{definition}
The orthogonal projection onto $T_{\mathbf{Y}}\widetilde{\mathcal{Y}}$ is a mapping $P_{T_{\mathbf{Y}}\widetilde{\mathcal{Y}}}:\mathbb{R}^{n\times n}\rightarrow T_{\mathbf{Y}}\widetilde{\mathcal{Y}}$ such that for a given matrix $\mathbf{A}_1\in\mathbb{R}^{n\times n}$, $<\mathbf{A}_1-P_{T_{\mathbf{Y}}\widetilde{\mathcal{Y}}}(\mathbf{A}_1),\mathbf{A}_2> = 0$ for all $\mathbf{A}_2\in T_{\mathbf{Y}}\widetilde{\mathcal{Y}}$.
\label{def:def002}
\end{definition} 

For a given matrix $\mathbf{A}_1$, orthogonal projection $P_{T_{\mathbf{Y}}\widetilde{\mathcal{Y}}}(\mathbf{A}_1)$ of $\mathbf{A}_1$ onto the tangent space $T_{\mathbf{Y}}\widetilde{\mathcal{Y}}$ is \cite{localization:bart2013}  
\begin{equation}
P_{T_{\mathbf{Y}}\widetilde{\mathcal{Y}}}(\mathbf{A}_1) = \mathbf{P}_\mathbf{Q}\text{Sym}(\mathbf{A}_1) + \text{Sym}(\mathbf{A}_1)\mathbf{P}_\mathbf{Q} - \mathbf{P}_\mathbf{Q}\text{Sym}(\mathbf{A}_1)\mathbf{P}_\mathbf{Q},
\label{eq:eq513}
\end{equation} 
where $\mathbf{P}_\mathbf{Q} = \mathbf{QQ}^T$.

\vspace{0.2cm}
In order to express the concept of moving in the direction of a tangent space while staying on the manifold, an operation called \textit{retraction} is used. As illustrated in Fig. \ref{fig:fig2}(b), the retraction operation is a mapping from $T_{\mathbf{Y}}\widetilde{\mathcal{Y}}$ to $\widetilde{\mathcal{Y}}$ that preserves the gradient at $\mathbf{Y}$ \cite[Definition 4.1.1]{localization:absil2012}. 
\begin{definition}
The retraction $R_{\mathbf{Y}}(\mathbf{B})$ of a matrix $\mathbf{B}\in T_{\mathbf{Y}}\widetilde{\mathcal{Y}}$ onto $\widetilde{\mathcal{Y}}$ is defined as    
\begin{equation}
R_{\mathbf{Y}}(\mathbf{B})=\arg\min\limits_{\mathbf{Z}\in\widetilde{\mathcal{Y}}}\|\mathbf{Y}+\mathbf{B}-\mathbf{Z}\|_F .
\label{eq:eq22}
\end{equation}
\end{definition}

In obtaining the closed form expression of $R_{\mathbf{Y}}(\mathbf{B})$, an operator $\mathcal{W}_k$ keeping $k$ largest positive eigenvalues of a matrix, referred to as \textit{eigenvalue selection operator}, is needed. Since the projection $R_{\mathbf{Y}}(\mathbf{B})$ is an element of $\widetilde{\mathcal{Y}}$, $R_{\mathbf{Y}}(\mathbf{B})$ should be a symmetric PSD matrix with rank $k$. Thus, for a given square matrix $\mathbf{A}$, we are interested only in the symmetric part $Sym(\mathbf{A})$. If we denote the eigenvalue decomposition (EVD) of this as $Sym(\mathbf{A}) = \mathbf{P}\boldsymbol\Sigma\mathbf{P}^T$ and the $k$ topmost eigenvalues of this as $\sigma_1\geq \sigma_2\geq \cdots \geq \sigma_k$ , then $\mathcal{W}_k(\mathbf{A})$ is defined as
\begin{equation}
\mathcal{W}_k(\mathbf{A}) = \mathbf{P}\boldsymbol\Sigma_k\mathbf{P}^T,
\label{eq:eq521}
\end{equation}
where $\boldsymbol\Sigma_k = \text{eye}\left(\matc{\sigma_1 & ... & \sigma_k &0& ...&0}^T\right)$. Using this eigenvalue selection operator $\mathcal{W}_k$, we can obtain an elegant expression of $R_{\mathbf{Y}}(\mathbf{B})$.  
\begin{theorem}[Proposition 6 \cite{localization:absil2012}]
The retraction $R_{\mathbf{Y}}(\mathbf{B})$ of $\mathbf{B}\in T_{\mathbf{Y}}\widetilde{\mathcal{Y}}$ can be expressed as 
\begin{equation}
R_\mathbf{Y}(\mathbf{B}) = \mathcal{W}_k(\mathbf{Y}+\mathbf{B}). 
\label{eq:eq554} 
\end{equation}   
\label{lm:lm005}
\end{theorem}

Finally, to develop the conjugate gradient algorithm over the Riemannian manifold $\widetilde{\mathcal{Y}}$, we need the Euclidean gradient of the cost function $f(\mathbf{Y})$.
\begin{theorem}
Euclidean gradient $\nabla_{\mathbf{Y}}f(\mathbf{Y})$ of $f(\mathbf{Y})$ with respect to $\mathbf{Y}$ is 
\begin{equation}
\nabla_{\mathbf{Y}}f(\mathbf{Y}) = 2\text{eye}(\text{Sym}(\mathbf{R})\mathbf{1})-2\text{Sym}(\mathbf{R}),
\label{eq:eq562}
\end{equation}
where $\mathbf{R}=\mathbf{W}\odot \mathbf{W}\odot (\mathcal{P}_E(g(\mathbf{Y})) - \mathcal{P}_E(\mathbf{D}_{o}))$.
\label{lm:lm006}  
\end{theorem}
\begin{proof}
See Appendix \ref{app:appB}.
\end{proof}

%----------------------------------------------------------
\begin{figure} [t] 
	\centering
  {\epsfig{figure=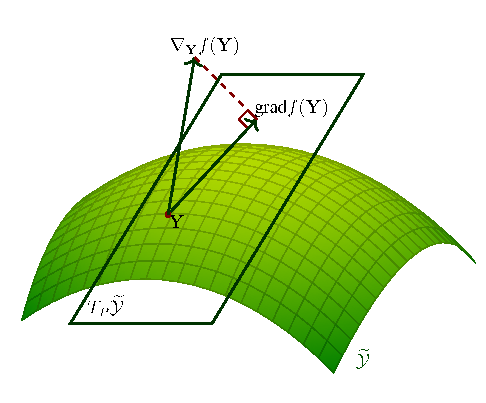, scale=1}}
  \caption {Riemannian gradient $\text{grad}f(\mathbf{Y})$ is defined as the projection of the Euclidean gradient $\nabla_{\mathbf{Y}}f(\mathbf{Y})$ onto the tangent space $T_{\mathbf{Y}}\widetilde{\mathcal{Y}}$ while the Euclidean gradient is a direction for which the cost function is reduced most in $\mathbb{R}^{n\times n}$, Riemannian gradient is the direction for which the cost function is reduced most in the tangent space $T_{\mathbf{Y}}\widetilde{\mathcal{Y}}$.}  
\label{fig:fig9}
\end{figure}
%----------------------------------------------------------

\subsection{Localization in Riemannian Manifold Using Conjugate Gradient (LRM-CG)}
\label{subsec:subsec002}
In order to solve the problem (\ref{eq:eq520}), we basically use the conjugate gradient (CG) algorithm. CG algorithm is widely used to solve the sparse symmetric positive definite linear systems \cite{localization:dai2011}. Main advantage of the CG algorithm is that the solution can be found in a finite number of searching steps. This is because the conjugate direction is designed such that it is conjugate to the previous directions and also the gradient of the cost function.     

First, noting that $\mathcal{P}_E$ and $g$ are linear mappings, one can easily show that
\begin{eqnarray}
f(\mathbf{Y}) & = & \frac{1}{2}\|\mathbf{W}\odot(\mathcal{P}_E(g(\mathbf{Y})) - \mathcal{P}_E(\mathbf{D}_{o}))\|_F^2\nonumber\\
 & = & \frac{1}{2}\|\mathbf{W}\odot(\mathcal{P}_E(g(\sum\limits_{i,j}y_{ij}\mathbf{e}_i\mathbf{e}_j^T))-\mathcal{P}_E(\mathbf{D}_{o}))\|_F^2 \nonumber\\
  & = & \frac{1}{2}\|\mathbf{W}\odot(\sum\limits_{i,j}y_{ij}\mathcal{P}_E(g(\mathbf{e}_i\mathbf{e}_j^T))-\mathcal{P}_E(\mathbf{D}_{o}))\|_F^2 \nonumber\\
& \substack{(a)\\ =} & \frac{1}{2}\|\text{vec}(\mathbf{W})\circ(\sum\limits_{i,j}y_{ij}\text{vec}\left(\mathcal{P}_E(g(\mathbf{e}_i\mathbf{e}_j^T))\right)\nonumber\\
& & - \text{vec}(\mathcal{P}_E(\mathbf{D}_{o})))\|_2^2 \nonumber \\
\label{eq:eq617}
& \substack{(b)\\ =} & \frac{1}{2}\|\mathbf{A}\text{vec}(\mathbf{Y})-\mathbf{b}\|_2^2,
\end{eqnarray}
where (a) is because $\|\mathbf{M}\|_F = \|\text{vec}(\mathbf{M})\|_2$, (b) follows from $\text{vec}(\mathbf{Y}) = \matc{y_{11} & y_{21} & \cdots & y_{nn}}^T$, $\mathbf{b} = \text{vec}(\mathbf{W}\odot\mathcal{P}_E(\mathbf{D}_{o}))$, and $\mathbf{A}$ formed by column vectors $\text{vec}\left(\mathbf{W}\odot\mathcal{P}_E(g(\mathbf{e}_i\mathbf{e}_j^T))\right)$.
  
In (\ref{eq:eq617}), we see that the cost function $f(\mathbf{Y})$ has the quadratic form of a sparse symmetric positive definite system, and thus the CG algorithm can be readily used to solve the problem. The update equation of the conventional CG algorithm in the Euclidean space is  
\begin{equation}
\mathbf{Y}_{i+1} = \mathbf{Y}_i + \alpha_i\mathbf{P}_i,
\label{eq:eq522}
\end{equation}  
where $\alpha_i$ is the stepsize and $\mathbf{P}_i$ is the conjugate direction. The stepsize $\alpha_i$ is chosen by the line minimization technique (e.g., Armijo's rule \cite{localization:dai2011}) and the search direction $\mathbf{P}_i$ of the CG algorithm is chosen as a linear combination of the gradient and the previous search direction to generate a direction conjugate to the previous ones. In doing so, one can avoid unnecessary searching of directions that have been searched over and thus achieve the speedup of the algorithm \cite{localization:dai2011}.

\begin{table*}[tp]
\centering
\begin{tabular}{l}
\hline 
\textbf{Algorithm 1:} LRM-CG algorithm \\
\hline
\begin{tabular}{rll}
\textbf{1} & \textbf{Input:} $\mathbf{D}_{o}$: the observed matrix, \\
& \quad\quad\:  $\mathbf{W}$: the weight matrix, \\
& \quad\quad\:  $\mathcal{P}_E$: the sampling operator, \\
& \quad\quad\:  $\epsilon$: tolerance, \\
& \quad\quad\:  $\mu\in (0\:\:\: 1)$: given constant, \\
& \quad\quad\:  $T$: number of iterations. & \\
\textbf{2} & \textbf{Output:}  $\widehat{\mathbf{X}}$: node location matrix &  \\
\textbf{3} & \textbf{Initialize:}  $i = 1$, & \\ 
& \quad\quad\:\:\:\: $\mathbf{Y}_1\in\widetilde{\mathcal{Y}}$: initial matrix, & \\
& \quad\quad\:\:\:\: $\mathbf{P}_1$: initial conjugate direction. & \\
\textbf{4} & \textbf{While}  $i\leq T$ \textbf{do} &\\
\textbf{5} & \quad $\mathbf{R}_i=\mathbf{W}\odot\mathbf{W}\odot(\mathcal{P}_E(g(\mathbf{Y}_i)) - \mathcal{P}_E(\mathbf{D}_{o}))$ & // Generate residual matrix \\
\textbf{6} & \quad  $\nabla_{\mathbf{Y}}f(\mathbf{Y}_i) = 2\text{eye}(\text{Sym}(\mathbf{R}_i)\mathbf{1})-2\mathbf{R}_i$ &  // Compute Euclidean gradient \\ 
\textbf{7} & \quad $\text{grad}f(\mathbf{Y}_i) = P_{T_{\mathbf{Y}_i}\widetilde{\mathcal{Y}}}(\nabla_{\mathbf{Y}} f(\mathbf{Y}_i))$ & // Compute Riemannian gradient \\
\textbf{8} & \quad $\mathbf{H}_i =  \text{grad}f(\mathbf{Y}_i) - P_{T_{\mathbf{Y}_i}\widetilde{\mathcal{Y}}}(\text{grad}f(\mathbf{Y}_{i-1}))$ & \\
\textbf{9} & \quad $h = <\mathbf{P}_i,\mathbf{H}_i>$ & \\
\textbf{10} & \quad $\beta_i = \frac{1}{h^2}<h\mathbf{H}_i-2\mathbf{P}_i\|\mathbf{H}_i\|_F^2,\text{grad}f(\mathbf{Y}_{i})>$ & // Compute CG coefficient \\
\textbf{11} & \quad $\mathbf{P}_i = -\text{grad}f(\mathbf{Y}_i) + \beta_iP_{T_{\mathbf{Y}_i}\widetilde{\mathcal{Y}}}(\mathbf{P}_{i-1})$ &  // Compute conjugate direction \\ 
\textbf{12} & \quad Find a stepsize $\alpha_i\textgreater 0$ such that &  // Perform Armijo's line search\\
& \quad\quad  $f(\mathbf{Y}_i) - f(R_{\mathbf{Y}_i}(\alpha_i\mathbf{P}_i))\geq -\mu\alpha_i<\text{grad}f(\mathbf{Y}_i),\mathbf{P}_i >$ &  \\
\textbf{13} & \quad  $\mathbf{Y}_{i+1}=R_{\mathbf{Y}_i}(\alpha_i\mathbf{P}_i)$ &  // Perform retraction \\
\textbf{14} & \quad $\mathbf{D}_{i+1} = g(\mathbf{Y}_{i+1})$ & // Compute updated Euclidean distance matrix \\
\textbf{15} & \quad \textbf{If}  $\|\mathbf{W}\odot(\mathcal{P}_E(\mathbf{D}_{i+1})-\mathcal{P}_E(\mathbf{D}_{o}))\|_F\textless \epsilon$ \textbf{then} &\\
\textbf{16} & \quad\quad Exit from while loop & \\
\textbf{17} & \quad \textbf{End If} & \\	
\textbf{18} & \quad Obtain $\mathbf{Q}$ and $\boldsymbol\Lambda$ using the eigendecomposition & \\
 & \quad\quad $\mathbf{Y}_{i+1} = \mathbf{Q}\boldsymbol\Lambda\mathbf{Q}^T$ & \\
\textbf{19} & \quad $\widehat{\mathbf{X}} = \mathbf{Q}\boldsymbol\Lambda^{1/2}$ &  // Find updated locations of sensor nodes \\
\textbf{20} & \quad $i=i+1$ & \\ 
\textbf{21} & \textbf{End While} & \\
\end{tabular}  \\
\hline
\end{tabular}
\end{table*}

Since we consider the optimization problem over the Riemannian manifold $\widetilde{\mathcal{Y}}$, the conjugate direction $\mathbf{P}_i$ should lie on the tangent space. To make sure that the update point $\mathbf{Y}_{i+1}$ lies on the manifold, we need a retraction operation. The update equation after applying the retraction operation is  
\begin{eqnarray}
\mathbf{Y}_{i+1} & = & R_{\mathbf{Y}_i}(\alpha_i\mathbf{P}_i) 
= \mathcal{W}_k(\mathbf{Y}_i+\alpha_i\mathbf{P}_i).
\label{eq:eq515}
\end{eqnarray} 
As observed in Theorem \ref{lm:lm005}, the eigenvalue selection operator $\mathcal{W}_k$ guarantees that the updated point $\mathbf{Y}_{i+1}$ lies on the manifold. 

We next consider the conjugate direction $\mathbf{P}_i$ of LRM-CG. In the conventional nonlinear CG algorithm, conjugate direction $\mathbf{P}_i$ is updated as
\begin{equation}
\mathbf{P}_i = - \nabla_{\mathbf{Y}}f(\mathbf{Y}_i) + \beta_i\mathbf{P}_{i-1},
\label{eq:eq518}
\end{equation} 
where $\beta_i$ is the conjugate update parameter\footnotemark\footnotetext{There are a number of ways to choose $\beta_i$. See, e.g., \cite{localization:dai2011,localization:hager2005}. }. Since we optimize over the Riemannian manifold $\widetilde{\mathcal{Y}}$, conjugate direction in (\ref{eq:eq518}) needs to be modified. First, we need to use the Riemannian gradient of $f(\mathbf{Y})$ instead of the Euclidean gradient $\nabla_{\mathbf{Y}}f(\mathbf{Y})$ since we need to find the search direction on the tangent space of $\widetilde{\mathcal{Y}}$. Riemannian gradient, denoted $\text{grad}f(\mathbf{Y})$, is distinct from $\nabla_{\mathbf{Y}}f(\mathbf{Y})$ in the sense that it is defined on the tangent space $T_{\mathbf{Y}}\widetilde{\mathcal{Y}}$ (see Fig. \ref{fig:fig9}). $\text{grad}f(\mathbf{Y})$ is given in the following lemma.
\begin{lemma}[Ch.3 \cite{localization:absil2008}]
Riemannian gradient $\nabla_{\mathbf{Y}}f(\mathbf{Y})$ of $f(\mathbf{Y})$ with respect to $\mathbf{Y}$ is
\begin{equation}
\text{grad}f(\mathbf{Y}) = P_{T_{\mathbf{Y}}\widetilde{\mathcal{Y}}}(\nabla_{\mathbf{Y}} f(\mathbf{Y})).
\label{eq:eq542}
\end{equation} 
\end{lemma}

Second, since the Riemannian gradient $\text{grad}f(\mathbf{Y}_i)$ and previous conjugate direction $\mathbf{P}_{i-1}$ lie on two different vector spaces $T_{\mathbf{Y}_i}\widetilde{\mathcal{Y}}$ and $T_{\mathbf{Y}_{i-1}}\widetilde{\mathcal{Y}}$, we need to project $\mathbf{P}_{i-1}$ onto the tangent space $T_{\mathbf{Y}_i}\widetilde{\mathcal{Y}}$ before performing a linear combination between of two\footnotemark\footnotetext{In transforming a component from one tangent space to another, an operator called vector transport is used (see Definition 8.1.1 in  \cite{localization:absil2008}). For an embedded manifold of $\mathbb{R}^{n\times n}$, vector transport is the orthogonal projection operator \cite{localization:absil2008}. Hence, the vector transport of $\mathbf{P}_{i-1}$ is the orthogonal projection of $\mathbf{P}_{i-1}$ onto $T_{\mathbf{Y}_i}\widetilde{\mathcal{Y}}$}. In view of this, the conjugate direction update equation of LRM-CG is
\begin{equation}
\mathbf{P}_i = - \text{grad}f(\mathbf{Y}_i) + \beta_iP_{T_{\mathbf{Y}_{i}}\widetilde{\mathcal{Y}}}(\mathbf{P}_{i-1}).
\label{eq:eq519}
\end{equation} 
In finding the stepsize $\alpha_i$ in (\ref{eq:eq515}), we use the Armijo's rule ($\alpha_i\approx\min\limits_{\alpha\textgreater 0}f(\mathcal{W}_k(\mathbf{Y}_i+\alpha_i\mathbf{P}_i)$ \cite{localization:dai2011}), a widely used line search strategy.

Finally, when the output $\widehat{\mathbf{Y}}\in\widetilde{\mathcal{Y}}$ of LRM-CG is generated, the node location matrix $\widehat{\mathbf{X}}$ is recovered as
\begin{equation}
\widehat{\mathbf{X}} = \arg\min\limits_{\mathbf{X}}\|\widehat{\mathbf{Y}} - \mathbf{XX}^T\|_F.
\label{eq:eq807}
\end{equation}  
Since $\widehat{\mathbf{Y}}\succeq 0$, we use the eigenvalue decomposition to find $\widehat{\mathbf{X}}$. In fact, by denoting $\widehat{\mathbf{Y}} = \mathbf{Q}\boldsymbol\Lambda\mathbf{Q}^T$ ($\mathbf{Q}\in\mathbb{R}^{n\times k}$ and $\boldsymbol\Lambda\in\mathbb{R}^{k\times k}$), we obtain the local locations of sensor nodes $\widehat{\mathbf{X}} = \mathbf{Q}\boldsymbol\Lambda^{1/2}$.
Then, $\widehat{\mathbf{X}}$ is transformed into the true locations of nodes by the aid of anchor nodes \cite{localization:shang2003,localization:rawat2014}. 
Specifically, we might need to find a rotation matrix $\mathbf{F}\in\mathbb{R}^{k\times k}$ and a translation vector $\mathbf{b}\in\mathbb{R}^{k}$ to transform $\widehat{\mathbf{X}}$ so that it would be matched with the true locations as 
\begin{equation}
\mathbf{X} = \widehat{\mathbf{X}}\mathbf{F} + \mathbf{1b}^T.
\end{equation} 
Suppose that $\mathbf{x}_1$, $\mathbf{x}_2$, and $\mathbf{x}_3$ are the given anchor nodes in 2-dimensional Euclidean space ($k = 2$). Since $\mathbf{x}_1^T = \widehat{\mathbf{x}}_1^T\mathbf{F} + \mathbf{b}^T$ and $\mathbf{b}^T = \mathbf{x}_1^T - \widehat{\mathbf{x}}_1^T\mathbf{F}$, we have 
\begin{align}
\mathbf{X} - \mathbf{1x}_1^T & = (\widehat{\mathbf{X}} - \mathbf{1}\widehat{\mathbf{x}}_1^T)\mathbf{F} + \mathbf{1}\widehat{\mathbf{x}}_1^T\mathbf{F} - \mathbf{1x}_1^T + \mathbf{1b}^T\notag\\
& = (\widehat{\mathbf{X}} - \mathbf{1}\widehat{\mathbf{x}}_1^T)\mathbf{F}.
\end{align}
What remains is to find $\mathbf{F}$. Let $\mathbf{X}_A = \matc{\mathbf{x}_2-\mathbf{x}_1 & \mathbf{x}_3 - \mathbf{x}_1}^T$ and $\widehat{\mathbf{X}}_A = \matc{\widehat{\mathbf{x}}_2-\widehat{\mathbf{x}}_1 & \widehat{\mathbf{x}}_3 - \widehat{\mathbf{x}}_1}$ be the matrices associated to anchor nodes. Then, it is clear that $\mathbf{X}_A = \widehat{\mathbf{X}}_A\mathbf{F}$. Thus, we have $\mathbf{F} = \widehat{\mathbf{X}}_A^\dagger\mathbf{X}_A$.
   
The proposed LRM-CG algorithm is summarized in Algorithm 1.

% Note that IEEE does not put floats in the very first column - or typically
% anywhere on the first page for that matter. Also, in-text middle ("here")
% positioning is not used. Most IEEE journals/conferences use top floats
% exclusively. Note that, LaTeX2e, unlike IEEE journals/conferences, places
% footnotes above bottom floats. This can be corrected via the \fnbelowfloat
% command of the stfloats package.

\subsection{Computational Complexity}
In this subsection, we analyze the computational complexity of LRM-CG in terms of the number of floating point operations (flops). As discussed, major operations in LRM-CG is to compute Euclidean gradient, Riemannian gradient, and the retraction operation.

First, in order to compute the Euclidean gradient $\nabla_{\mathbf{Y}}f(\mathbf{Y}_i)$ in (\ref{eq:eq562}), we need to compute $\mathcal{P}_E(g(\mathbf{Y}_i))$, $\mathbf{R}_i =\mathbf{W}\odot \mathbf{W}\odot (\mathcal{P}_E(g(\mathbf{Y}_i)) - \mathcal{P}_E(\mathbf{D}_{o}))$, and $2\text{eye}(\text{Sym}(\mathbf{R}_i)\mathbf{1})-2\text{Sym}(\mathbf{R}_i)$, which require $3|E|/2$, $3|E|/2$, and $\mathcal{O}(|E|)$ flops\footnote{To compute $2\text{eye}(\text{Sym}(\mathbf{R}_i)\mathbf{1})-2\text{Sym}(\mathbf{R}_i)$, we need $\text{Sym}(\mathbf{R}_i)$ and $\text{eye}(\text{Sym}(\mathbf{R}_i)\mathbf{1})$ given $\text{Sym}(\mathbf{R}_i)$, which requires $|E|$ and $\mathcal{O}(|E|)$ flops, respectively.}, respectively, where $|E|$ is the number of elements of $E$ defined in \eqref{eq:eq209}.     
Second, recalling that the Riemannian gradient $\text{grad}f(\mathbf{Y}_i)$ is an orthogonal projection of $\nabla_{\mathbf{Y}}f(\mathbf{Y}_i)$ onto the tangent space $T_{\mathbf{Y}_i}\widetilde{\mathcal{Y}}$, we need to estimate the computational complexity of the orthogonal projection operator $P_{T_{\mathbf{Y}_i}\widetilde{\mathcal{Y}}}$. By adopting the computational strategy in \cite{localization:bart2013}, we can express the Riemannian gradient as $\text{grad}f(\mathbf{Y}_i) = \mathbf{Q}\mathbf{C}_1\mathbf{Q}^T+\mathbf{Q}\widetilde{\mathbf{C}}^T+\widetilde{\mathbf{C}}\mathbf{Q}^T$ where $\mathbf{C}_1\in\mathbb{R}^{k\times k}$, $\widetilde{\mathbf{C}}\in\mathbb{R}^{n\times k}$, and $\mathbf{Q}^T\widetilde{\mathbf{C}} = \mathbf{0}$, which requires $\mathcal{O}(k|E|+k^2n)$.     
Finally, the retraction operation is obtained via the eigenvalue selection operator $\mathcal{W}_k$ in (\ref{eq:eq554}) and this requires the EVD of $(\mathbf{Y}_i+\mathbf{P}_i)$. In general, the computational complexity of the EVD for an $n\times n$ matrix is expressed as $\mathcal{O}(n^3)$. 
However, by exploiting the symmetric structures of $\widetilde{\mathcal{Y}}$ and $T_{\mathbf{Y}_i}\widetilde{\mathcal{Y}}$, we can simplify the EVD operation (see \cite{localization:bart2013} for details). Specifically, noting that $\mathbf{P}_i\in T_{\mathbf{Y}_i}\widetilde{\mathcal{Y}}$, we have
\begin{eqnarray}
\mathbf{Y}_i + \mathbf{P}_i & = & \left[\begin{matrix}
\mathbf{Q} & \mathbf{Q}_{c}
\end{matrix} \right]\left[\begin{matrix}
\boldsymbol\Lambda+\mathbf{C}_{1} & \mathbf{R}_c^T\\
\mathbf{R}_c & \mathbf{0}
\end{matrix} \right]\left[\begin{matrix}
\mathbf{Q}^T\\
\mathbf{Q}_{c}^T
\end{matrix} \right] \nonumber\\
& = & \left[\begin{matrix}
\mathbf{Q} & \mathbf{Q}_{c}
\end{matrix} \right]\textbf{K}\boldsymbol\Lambda^{'}\textbf{K}^T\left[\begin{matrix}
\mathbf{Q}^T\\
\mathbf{Q}_{c}^T
\end{matrix} \right] \nonumber .
\end{eqnarray}
where $\mathbf{Q}_c\mathbf{R}_c$ is the QR-decomposition of an $n\times k$ matrix satisfying $\mathbf{Q}^T\mathbf{Q}_c = \mathbf{0}$, which requires $\mathcal{O}(k^2n + k^3)$. Now the EVD of $(\mathbf{Y}_i + \mathbf{P}_i)$ is simplified to the EVD of the $2k\times 2k$ matrix $\left[\begin{matrix}
\boldsymbol\Lambda + \mathbf{C}_1 & \mathbf{R}_c^T\\
\mathbf{R}_c & \mathbf{0}
\end{matrix} \right]$, which requires only $\mathcal{O}(k^3)$ flops. Also, the computation of $\left[\begin{matrix}
\mathbf{Q} & \mathbf{Q}_{c}
\end{matrix} \right]\textbf{K}$ requires $nk(4k-1)$ flops\footnote{Since $\mathbf{K}\in\mathbb{R}^{2k\times k}$ and $\left[\begin{matrix}
\mathbf{Q} & \mathbf{Q}_{c}
\end{matrix} \right]\in\mathbb{R}^{n\times 2k}$, each entry of $\left[\begin{matrix}
\mathbf{Q} & \mathbf{Q}_{c}
\end{matrix} \right]\textbf{K}$ requires $2k$ multiplications and $2k-1$ additions.}. 
As a result, the computational complexity of the retraction operation is $\mathcal{O}(k^2n)$.

In summary, the computational complexity of the proposed algorithm per iteration is $\mathcal{O}(k|E|+k^2n)$. Since $k = 2$ or $3$ in our problem \cite{localization:dattorro2005}, the computational complexity per iteration can be expressed as $ \mathcal{O}(|E| + n)$ flops.

\section{Convergence Analysis}
\label{sec:sec3}
In this section, we present the convergence analysis of the proposed LRM-CG algorithm. We show that under the extended Wolfe's conditions, LRM-CG converges linearly to the original Euclidean distance matrix in the sampling space.

\vspace{0.2cm}
\begin{definition}
For a sequence of matrices $\{\mathbf{D}_i\}_{i=1}^\infty$, if $\lim\limits_{i\rightarrow\infty}\|\mathbf{D}_i-\mathbf{D}\|_F = 0$, we say $\{\mathbf{D}_i\}_{i=1}^\infty$ converges to $\mathbf{D}$. Further, we say $\{\mathbf{D}_i\}_{i=1}^\infty$ converges linearly to $\mathbf{D}$ with convergent rate $\lambda$ if there exists $\lambda$ ($1\textgreater \lambda\geq 0$) satisfying
\begin{equation}
\lim\limits_{i\rightarrow\infty}\quad \frac{\|\mathbf{D}_{i+1}-\mathbf{D}\|_F}{\|\mathbf{D}_{i}-\mathbf{D}\|_F} = \lambda. \nonumber
\end{equation} 
\end{definition}

The minimal set of assumptions used for the analytical tractability are as follows:
\begin{itemize}
\item[$\mathbf{A1}:$] $f(\mathbf{Y}_i) - f(R_{\mathbf{Y}_i}(\alpha_i\mathbf{P}_i))  \geq  -\tau \alpha_i<\text{grad}f(\mathbf{Y}_i),\mathbf{P}_i >$ for $\tau$ satisfying $0\textless \tau \textless 1/2$, 
\item[$\mathbf{A2}:$] $|<\text{grad}f(R_{\mathbf{Y}_i}(\alpha_i\mathbf{P}_i)),\mathbf{P}_i >| \leq  -\mu <\text{grad}f(\mathbf{Y}_i),\mathbf{P}_i >$ for $\mu$ satisfying $\tau\textless \mu\textless 1/2$,
\item[$\mathbf{A3}:$] $c\|\text{grad}f(\mathbf{Y}_i)\|_F \geq  \|\nabla_{\mathbf{Y}}f(\mathbf{Y}_{i})\|_F$ for $c$ satisfying $c\textgreater 1$. 
\end{itemize}

In essence, $\mathbf{A1}$ and $\mathbf{A2}$ can be considered as extensions of the strong Wolfe's conditions \cite{localization:wolfe1969,localization:sato2015}. 
Note that if the stepsize $\alpha_i$ is chosen to be very small, then $\mathbf{Y}_{i+1} = R_{\mathbf{Y}_i}(\alpha_i\mathbf{P}_i) \approx R_{\mathbf{Y}_i}(\mathbf{0}) = \mathbf{Y}_i$, and thus $ f(\mathbf{Y}_i) - f(R_{\mathbf{Y}_i}(\alpha_i\mathbf{P}_i))  \approx   0$ and $-\tau \alpha_i<\text{grad}f(\mathbf{Y}_i),\mathbf{P}_i >  \approx  0$. In this case, \textbf{A1} holds true approximately. However, there would be almost no update of $\mathbf{Y}_i$ so that the algorithm will converge extremely slowly. To circumvent this pathological scenario, we use \textbf{A2}, which is in essence an extension of the strong Wolfe's condition for the Riemannian manifold. Under this assumption, $\alpha_i$ cannot be chosen to be very small since otherwise we have $R_{\mathbf{Y}_i}(\alpha_i\mathbf{P}_i) \approx \mathbf{Y}_i$, and thus  
\begin{align}
|<\text{grad}f(R_{\mathbf{Y}_i}(\alpha_i\mathbf{P}_i)),\mathbf{P}_i >|  & \approx  |<\text{grad}f(\mathbf{Y}_i),\mathbf{P}_i >|  \notag\\
& \geq  -\mu <\text{grad}f(\mathbf{Y}_i),\mathbf{P}_i >, \notag
\end{align}  
which contradicts the assumption \textbf{A2}.  
The assumption \textbf{A3} is needed to guarantee the global convergence of LRM-CG. We will discuss more on this in Remark \ref{rmk:rmk002}.

Our first main result, stating successful recovery condition at sampled entries, is formally described in the following theorem.
\begin{theorem}[strong convergence of LRM-CG]
Let $\{\mathbf{D}_i = g(\mathbf{Y}_i)\}_{i=1}^\infty$ be the sequence of the matrices generated by LRM-CG and $\mathbf{D}$ be the original Euclidean distance matrix. Under $\mathbf{A1}$, $\mathbf{A2}$, and $\mathbf{A3}$, $\{\mathcal{P}_E(\mathbf{D}_i)\}_{i=0}^\infty$ converges linearly to $\mathcal{P}_E(\mathbf{D})$.
\label{thm:thm001} 
\end{theorem}

\vspace{0.2cm}

\begin{remark}[strongly convergent condition in $\mathbb{R}^n$]
\label{rmk:rmk001}
Note that $\lim\limits_{i\rightarrow\infty}\|\mathcal{P}_E(\mathbf{D}_i)-\mathcal{P}_E(\mathbf{D})\|_F = 0$ is equivalent to
\begin{equation}
\lim\limits_{i\rightarrow \infty} \|\nabla_\mathbf{Y}f(\mathbf{Y}_i)\|_F = 0.
\label{eq:eq707}
\end{equation}
This condition is often referred to as the \textit{strongly convergent condition} of the nonlinear CG algorithms in the vector space. The equivalence can be established by the following sandwich lemma. 
\begin{lemma}
\begin{align}
2\|\mathcal{P}_E(\mathbf{D}_i)-\mathcal{P}_E(\mathbf{D})\|_F & \leq \|\nabla_\mathbf{Y}f(\mathbf{Y}_i)\|_F \notag\\
& \leq (2\sqrt{n}+2)\|\mathcal{P}_E(\mathbf{D}_i)-\mathcal{P}_E(\mathbf{D})\|_F.\notag
\end{align}
\label{lm:lm308}
\end{lemma}
\vspace{-0.5cm}
\begin{proof}
See Appendix \ref{app:appAJ}
\end{proof}
\end{remark}

%----------------------------------------------------------
\begin{figure} [t] 
	\centering
  {\epsfig{figure=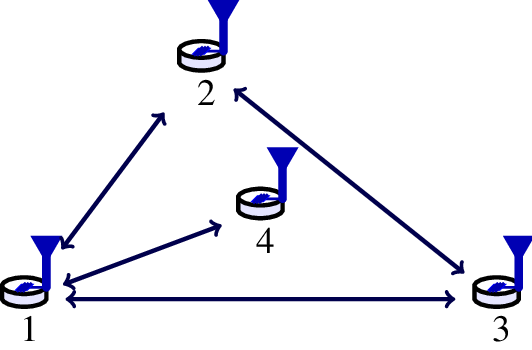, scale=0.95}}
  \caption {Suppose that the sensor node 4 is inside the triangle formed by three sensor nodes 1, 2, and 3. Then for a given $r$, it can be shown that $d_{14}\leq \max(d_{12},d_{13})$, and thus $P(d_{14}\leq r|d_{12}\leq r, d_{13}\leq r) = 1$ which is not necessarily equivalent to $P(d_{14}\leq r)$.}  
\label{fig:fig002}
\end{figure}
%----------------------------------------------------------

\begin{remark}
\label{rmk:rmk002}
Recently, an attempt has been made to extend the convergent analysis of the conventional CG algorithms (over the Euclidean space $\mathbb{R}^n$) to the Riemannian manifolds. In  \cite[Theorem 4.3]{localization:sato2015}, it has been shown that under certain assumption,
\begin{equation}
\lim\limits_{i\rightarrow \infty}\inf \|\text{grad}f(\mathbf{Y}_i)\|_F = 0.
\label{eq:eq708}
\end{equation}
One can observe that the Euclidean gradient $\nabla_\mathbf{Y}f(\mathbf{Y}_i)$ is replaced by the Riemannian gradient $\text{grad}f(\mathbf{Y}_i)$. Unfortunately, the convergence of the Riemannian gradient in (\ref{eq:eq708}) does not imply the convergence of Euclidean gradient in (\ref{eq:eq707}) because
\begin{eqnarray}
& & \|\nabla_\mathbf{Y}f(\mathbf{Y}_i)\|_F^2 \nonumber\\
& = & \|\mathcal{P}_{T_{\mathbf{Y}}\widetilde{\mathcal{Y}}}(\nabla_\mathbf{Y}f(\mathbf{Y}_i))\|_F^2 + \|\mathcal{P}^\perp_{T_{\mathbf{Y}}\widetilde{\mathcal{Y}}}(\nabla_\mathbf{Y}f(\mathbf{Y}_i))\|_F^2\nonumber \\
& = & \|\text{grad}f(\mathbf{Y}_i)\|_F^2 + \|\mathcal{P}^\perp_{T_{\mathbf{Y}}\widetilde{\mathcal{Y}}}(\nabla_\mathbf{Y}f(\mathbf{Y}_i))\|_F^2, 
\label{eq:eq709}
\end{eqnarray}
where $\text{grad}f(\mathbf{Y}_i) = \mathcal{P}_{T_{\mathbf{Y}}\widetilde{\mathcal{Y}}}(\nabla_\mathbf{Y}f(\mathbf{Y}_i))$ (see (\ref{eq:eq542})). One can observe from this that the condition in (\ref{eq:eq708}) is not sufficient to guarantee (\ref{eq:eq707}), that is, one cannot guarantee $\lim\limits_{i\rightarrow\infty}\|\mathcal{P}_E(\mathbf{D}_i)-\mathcal{P}_E(\mathbf{D})\|_F = 0$ just from (\ref{eq:eq708}). However, by the introduction of \textbf{A3}, equivalence between (\ref{eq:eq707}) and (\ref{eq:eq708}) can be established.

\end{remark}

We are now ready to prove Theorem \ref{thm:thm001}.

\subsubsection*{Proof of Theorem \ref{thm:thm001}}

First, we show that under \textbf{A1}, \textbf{A2}, and \textbf{A3}, $\|\mathcal{P}_E(\mathbf{D}_{i}) - \mathcal{P}_E(\mathbf{D})\|_F  $ is non-increasing. 
That is, there exists $\gamma\textgreater 0$ such that $\frac{\gamma}{c^2}\leq 1$ and
\begin{equation}
\|\mathcal{P}_E(\mathbf{D}_{i+1}) - \mathcal{P}_E(\mathbf{D})\|_F^2 \leq \left(1 - \frac{\gamma}{c^2}\right)\|\mathcal{P}_E(\mathbf{D}_{i}) - \mathcal{P}_E(\mathbf{D})\|_F^2.
\label{eq:eq703}
\end{equation}
We need the following lemma to prove this.

\begin{lemma}
Suppose that $\|\text{grad}f(\mathbf{Y}_{t})\|_F>0$ for all $t\leq i$. If $\beta_i$ is chosen based on Fletcher-Reeves' rule, that is, \cite{localization:hager2005,localization:sato2015}    
\begin{equation}
\beta_i = \frac{<\text{grad}f(\mathbf{Y}_{i}),\text{grad}f(\mathbf{Y}_{i})>}{<\text{grad}f(\mathbf{Y}_{i-1}),\text{grad}f(\mathbf{Y}_{i-1})>},
\label{eq:eq716}
\end{equation}   
then
\begin{equation}
\frac{<\text{grad}f(\mathbf{Y}_{i}),\mathbf{P}_{i} >}{\|\text{grad}f(\mathbf{Y}_{i})\|_F^2}\leq - \frac{1-2\mu}{1-\mu} - \frac{\mu^{i}}{1-\mu} \nonumber .
\end{equation}
\label{lm:lm300}
\end{lemma}

\begin{lemma}  
$\|\text{grad}f(\mathbf{Y}_i)\|_F^2\geq \frac{8}{c^2}f(\mathbf{Y}_i).$
\label{lm:lm301}
\end{lemma}
\begin{proof}
See Appendix \ref{app:appAA}.
\end{proof} 

\vspace{0.5cm}

We are now ready to prove (\ref{eq:eq703}). First, from $\mathbf{A1}$, we have
\begin{align}
f(\mathbf{Y}_{i+1}) & \leq  f(\mathbf{Y}_i) + \tau\alpha_i<\text{grad}f(\mathbf{Y}_i),\mathbf{P}_i>\notag\\
& \substack{(a)\\ \leq}  f(\mathbf{Y}_i) - \tau\alpha_i\left(\frac{1-2\mu}{1-\mu}+\frac{\mu^i}{1-\mu} \right)\|\text{grad}f(\mathbf{Y}_i)\|_F^2 \notag \\
& \leq  f(\mathbf{Y}_i) - \tau\alpha_i\left(\frac{1-2\mu}{1-\mu} \right)\|\text{grad}f(\mathbf{Y}_i)\|_F^2,\notag\\
& \substack{(b)\\ \leq}  f(\mathbf{Y}_i) - 8\tau\alpha_i\left(\frac{1-2\mu}{1-\mu} \right)\frac{1}{c^2} f(\mathbf{Y}_i),\notag
\end{align}
where (a) and (b) follow from Lemma \ref{lm:lm300} and Lemma \ref{lm:lm301}, respectively. Let 
\begin{equation}
\gamma_i = 8\tau\alpha_i\left(\frac{1-2\mu}{1-\mu} \right),
\label{eq:eq913}
\end{equation}
then $\gamma_i\textgreater 0$ (since $\alpha_i\textgreater 0$) and hence 
\begin{equation}
f(\mathbf{Y}_{i+1})  \leq (1 - \frac{\gamma_i}{c^2})f(\mathbf{Y}_i). 
\label{eq:eq912}
\end{equation}
Recalling that $f(\mathbf{Y}_i) = \frac{1}{2}\|\mathcal{P}_E(\mathbf{D}_i) - \mathcal{P}_E(\mathbf{D})\|_F^2$, we have $$\|\mathcal{P}_E(\mathbf{D}_{i+1}) - \mathcal{P}_E(\mathbf{D})\|_F^2 \leq \left(1 - \frac{\gamma_i}{c^2}\right)\|\mathcal{P}_E(\mathbf{D}_{i}) - \mathcal{P}_E(\mathbf{D})\|_F^2.$$ By choosing $\gamma = \min\limits_i \gamma_i$, we get the desired result.

Now, what remains is to show that $\lim\limits_{i\rightarrow\infty} \|\mathcal{P}_E(\mathbf{D}_{i}) - \mathcal{P}_E(\mathbf{D})\|_F = 0$ under (\ref{eq:eq703}). Noting that $c > 1$, one can easily show that $1\textgreater (1-\frac{\gamma}{c^2})^{1/2}$. Using this together with (\ref{eq:eq703}), we have
\begin{eqnarray}
& & \lim\limits_{i\rightarrow\infty} \frac{\|\mathcal{P}_E(\mathbf{D}_{i+1}) - \mathcal{P}_E(\mathbf{D})\|_F}{\|\mathcal{P}_E(\mathbf{D}_{i}) - \mathcal{P}_E(\mathbf{D})\|_F} = (1-\frac{\gamma}{c^2})^{1/2}\textless 1 \nonumber\\
& & \text{and hence}\nonumber\\
& & \lim\limits_{i\rightarrow\infty}\|\mathcal{P}_E(\mathbf{D}_{i}) - \mathcal{P}_E(\mathbf{D})\|_F = 0 \nonumber .
\end{eqnarray}   
Thus, the sequence $\{\mathcal{P}_E(\mathbf{D}_{i})\}_{i=1}^\infty$ converges linearly to $\mathcal{P}_E(\mathbf{D})$.

\begin{remark}
\label{rmk:rmk003}
The global convergence of LRM-CG with a linear convergence rate is established based on \textbf{A3}. It would be appealing to relate \textbf{A3} to the problem parameters such as the radio communication range $r$ and the number of the observed entries $|E|$. In many studies, an assumption that observed entries of a desired low-rank matrix are sampled independently is commonly used \cite{localization:candes2009,localization:cai2010}. Unfortunately, this assumption cannot be used for the Euclidean distance matrix completion since the pairwise distances might not be sampled independently. For example, consider the scenario illustrated in Fig. \ref{fig:fig002}. Since the sensor node 4 is located inside the triangle formed by three sensor nodes (nodes 1, 2, and 3), one can see that $d_{14}\leq \max(d_{12},d_{13})$. Thus, if $d_{12}$ and $d_{13}$ are already observed (i.e., $d_{12}\leq r$, $d_{13}\leq r$), then so is $d_{14}$. In other words, $P(d_{14}\leq r|d_{12}\leq r, d_{13}\leq r) = 1$, while $P(d_{14}\leq r)$ is not necessarily one. This makes it difficult to connect \textbf{A3} and the problem parameters. To work around this issue, we introduce a positive constant $\epsilon$ and the $\epsilon$-relaxed version of \textbf{A3}: 
\begin{equation}
c^2\|\text{grad}f(\mathbf{Y}_i)\|_F^2 + \epsilon \textgreater \|\nabla_{\mathbf{Y}}f(\mathbf{Y}_i)\|_F^2,
\label{eq:eq910}
\end{equation}
which holds true with the probability proportional to $r$ and $|E|$. It is clear that \eqref{eq:eq910} is weaker than \textbf{A3}. Using \eqref{eq:eq910} instead of \textbf{A3}, it can be shown that the proposed LRM-CG algorithm converges locally with a linear convergence rate.     
\end{remark}

\section{Outlier Problem}
\label{sec:sec5}
In many practical scenarios, observed pairwise distances can be contaminated by the outliers. The outliers occur due to the power outage, obstacles, adversary attacks, or hardware (Tx/Rx) malfunction. Put it rigorously, an entry $d^{o}_{ij}$ of the observed matrix $\mathbf{D}_{o}$ is called an outlier if $d^{o}_{ij}\neq d_{ij}$ \cite{localization:yang2013}. Often we use the relaxed definition using the tolerance level $\rho$ of observation error. That is, $d^{o}_{ij}$ is defined as an outlier if $|d^{o}_{ij} - d_{ij}|\textgreater \rho$. Since the outlier often degrades the localization performance severely, we should control it in the recovery process. 

First, we model the observed distance as $d^{o}_{ij} = d_{ij} + l_{ij}$ ($l_{ij}$ is the outlier). Thus, $\mathcal{P}_E(\mathbf{D}_{o}) = \mathcal{P}_E(\mathbf{D} + \mathbf{L})$ where $\mathbf{L}$ is the outlier matrix. Since $\mathbf{L}$ is considered as a sparse matrix, we can modify the problem in (\ref{eq:eq520}) as
\begin{eqnarray}
& \min\limits_{\substack{\mathbf{Y}\:\in\:\widetilde{\mathcal{Y}} \\  \mathbf{L}\in\mathbb{R}^{n\times n}}} & \frac{1}{2}\|\mathbf{W}\odot(\mathcal{P}_E(g(\mathbf{Y})) + \mathcal{P}_E(\mathbf{L}) -\mathcal{P}_E(\mathbf{D}_{o}))\|_F^2\nonumber\\
& &  + \tau\|\mathbf{L}\|_o, 
\label{eq:eq801}
\end{eqnarray}
where $\|\mathbf{L}\|_o$ is the number of nonzero entries of $\mathbf{L}$ and $\tau$ is the regularization factor controlling the tradeoff between the sparsity of $\mathbf{L}$ and the consistency of the observed distances. 
Since $\|\mathbf{L}\|_o$ is nonlinear and non-convex, we instead use the convex surrogate $\|\mathbf{L}\|_{1} = \sum\limits_{i=1}^n\sum\limits_{j=1}^n |l_{ij}|$, and thus
\begin{eqnarray}
& \min\limits_{\substack{\mathbf{Y}\:\in\:\widetilde{\mathcal{Y}} \\  \mathbf{L}\in\mathbb{R}^{n\times n}}} & \frac{1}{2}\|\mathbf{W}\odot(\mathcal{P}_E(g(\mathbf{Y})) + \mathcal{P}_E(\mathbf{L}) -\mathcal{P}_E(\mathbf{D}_{o}))\|_F^2\nonumber\\
& & + \tau\|\mathbf{L}\|_1.
\label{eq:eq802}
\end{eqnarray}

Second, we use a slight modification version of the proposed LRM-CG and a soft-thresholding operator to find the solutions $\mathbf{Y}$ and $\mathbf{L}$ of (\ref{eq:eq802}), respectively. Specifically, the problem in (\ref{eq:eq802}) can be solved iteratively using alternative minimization as  
\begin{eqnarray}
\mathbf{Y}_{i+1} & = & \arg\min\limits_{\mathbf{Y}\:\in\:\widetilde{\mathcal{Y}}}  \frac{1}{2}\|\mathbf{W}\odot(\mathcal{P}_E(g(\mathbf{Y})) + \mathcal{P}_E(\mathbf{L}_i)\nonumber\\
\label{eq:eq803}
& & -\mathcal{P}_E(\mathbf{D}_{o}))\|_F^2 + \tau\|\mathbf{L}_i\|_1 \\
\mathbf{L}_{i+1} & = & \arg\min\limits_{\mathbf{L}\in\mathbb{R}^{n\times n}}  \frac{1}{2}\|\mathbf{W}\odot(\mathcal{P}_E(g(\mathbf{Y}_{i+1})) + \mathcal{P}_E(\mathbf{L})\nonumber\\
\label{eq:eq804}
& & -\mathcal{P}_E(\mathbf{D}_{o}))\|_F^2 + \tau\|\mathbf{L}\|_1 .
\end{eqnarray}
The subproblem in (\ref{eq:eq803}) can be solved using the proposed LRM-CG with simple modifications of the cost function and the residual matrix $\mathbf{R}_i$ in Algorithm 1. The modified residual is
\begin{equation}
\mathbf{R}_i = \mathbf{W}\odot \mathbf{W}\odot(\mathcal{P}_E(g(\mathbf{Y}_i)) + \mathcal{P}_E(\mathbf{L}_i) - \mathcal{P}_E(\mathbf{D}_{o})).
\label{eq:eq806}
\end{equation}  
Note that $\mathcal{P}_E(\mathbf{L}_i)$ is added to the original residual $\mathbf{R}_i$. The subproblem in (\ref{eq:eq804}) can be solved using the soft-thresholding operator, which gradually truncates the magnitude of the entries of a matrix \cite{localization:lin2010}. For a given matrix $\mathbf{A}$, the soft-thresholding operator output $\mathcal{T}(\mathbf{A})$ is defined as 
\begin{equation}
\mathcal{T}(a_{ij}) = \left\lbrace\begin{matrix}
\frac{w_{ij}a_{ij}-\tau}{w_{ij}^2} & \text{if } w_{ij}a_{ij}\geq \tau\\
\frac{w_{ij}a_{ij}+\tau}{w_{ij}^2} & \text{if } w_{ij}a_{ij}\leq - \tau\\
0 & \text{else}
\end{matrix}  \right. .\nonumber
\end{equation}
Using the soft-thresholding operator, the solution of (\ref{eq:eq804}) is given by \cite{localization:lin2010} 
\begin{equation}
\mathbf{L}_{i+1} = \mathcal{T}(\mathbf{W}\odot(\mathcal{P}_E(\mathbf{D}_{o}) - \mathcal{P}_E(g(\mathbf{Y}_{i+1})))).
\label{eq:eq805}
\end{equation}
In the sequel, we call this modified version of LRM-CG as the extended LRM-CG (ELRM-CG)\footnote{By extending the convergence analysis of LRM-CG, we can readily obtain the convergence guarantee of ELRM-CG. First, for the subproblem \eqref{eq:eq803}, we can trivially extend the convergence analysis of the problem \eqref{eq:eq520} in Section \ref{sec:sec3} and then have $h(\mathbf{Y}_{i+1}, \mathbf{L}_{i})\leq h(\mathbf{Y}_{i}, \mathbf{L}_{i})$ where $h$ is the cost function of \eqref{eq:eq802}. Second, for the subproblem \eqref{eq:eq804}, we can compute $\mathbf{L}_{i+1}$ in one step using the soft-thresholding operator in \eqref{eq:eq805} and thus we always have $h(\mathbf{Y}_{i+1}, \mathbf{L}_{i+1}) \leq h(\mathbf{Y}_{i+1}, \mathbf{L}_{i})$. Combining these, we have $h(\mathbf{Y}_{i+1}, \mathbf{L}_{i+1}) \leq h(\mathbf{Y}_{i+1}, \mathbf{L}_{i}) \leq h(\mathbf{Y}_{i}, \mathbf{L}_{i})$ for all $i$, which ensures the convergence of ELRM-CG. }.

\section{Simulation Results and Discussion}
\label{sec:sec4}

In this subsection, we test the performance of the proposed LRM-CG. In our simulations, we compare LRM-CG with following matrix completion algorithms:
\begin{itemize}
\item APG \cite{localization:lin2009}: an algorithm to solve the robust PCA problem via an accelerated proximal gradient method.
\item LRGeomCG \cite{localization:bart2013}: this algorithm can be considered as the CG algorithm defined over the Riemannian manifold of low rank matrices (but not necessarily positive definite).
\item SVT \cite{localization:cai2010}: an algorithm to solve the NNM problem using a singular value thresholding technique.
\item TNNR-ADMM \cite{localization:hu2013}: an algorithm to solve the truncated NNM problem via an alternating direction method of multipliers.   
\end{itemize} 
Also, we compare LRM-CG with the following localization algorithms:
\begin{itemize}
\item MDS \cite{localization:shang2003}: this is a multiscaling dimensional algorithm based on the shortest path algorithm and truncated eigendecomposition.
\item SDP \cite{localization:biswas2004, localization:guo2016}: an algorithm to solve the localization problem using a convex relaxation of nonconvex quadratic constraints of the node locations.  
\end{itemize}

In the scenario without observation error, we generate an $n\times k$ location matrix $\mathbf{X}$ whose entries are sampled independently and identically from a uniform distribution in the interval with 50 meters. Then, $\mathbf{X}$ is mapped into the Euclidean distance matrix $\mathbf{D} = g(\mathbf{XX}^T)$. As mentioned, an entry $d_{ij}^{o}$ of $\mathbf{D}_{o}$ is known (observed) if it is smaller than the radio communication range (i.e., $d_{ij}^{o}\leq r$). In the scenario with observation error, an observation error matrix $\mathbf{N}\in\mathbb{R}^{n\times n}$ is added to $\mathbf{D}$. In general, the accuracy of the observed distances is inversely proportional to the true distances \cite{localization:costa2006, localization:jianwu2009}. In our simulations, we employ the RSS-based model in which the cumulative effect of many attenuation factors of the wireless communication environment results in a log-normal distribution of the received power \cite{localization:costa2006}. Specifically, let $\delta$ be a normal random variable with zero mean and variance $\sigma_{dB}^2$. Then, each entry $n_{ij}$ of $\mathbf{N}$ is $n_{ij} = (\kappa 10^{\frac{\delta}{10n_p}} - 1)d_{ij}$ where $\delta$ is the constant dB error in the received power measurement, $n_p$ is the path loss parameter, and $\kappa = 10^{-\frac{\sigma_{dB}^2\ln 10}{200n_p^2}}$ is a constant to enforce the unbiasedness of the observed distances (i.e., $E[n_{ij}] = 0$).
In measuring the performance for each algorithm, we perform at least 1000 independent trials.

In the proposed LRM-CG, we use a random initialization in which the initial entries of $\mathbf{X}$ and $\mathbf{L}$ are chosen from i.i.d. standard normal random variables. 
In the simulation with observation errors, we choose the weight matrix to suppress the large magnitude errors. For the $(i,j)$-th entry $w_{ij}$ of $\mathbf{W}$ (see (\ref{eq:eq511})), we consider two settings. To account for the RSS-based measurement model, we set $w_{ij}$ inversely proportional to the error term $|d^{o}_{ij} - d_{ij}|$ as 
\begin{equation}
w_{ij} = w_{ij}^\ast = \left\lbrace \begin{matrix}
\exp(-|d^{o}_{ij} - \widetilde{d}_{ij}|^{\frac{1}{4}}) & \text{if } (i,j)\in E\\
0 & \text{else}
\end{matrix} \right. ,
\label{eq:eq808} 
\end{equation}
where $\widetilde{d}_{ij} = d_{ij}^{o}c^{3/4}/(1+\sqrt{c^{1/8}-1})^4$ is an estimate of $d_{ij}$\footnote{Using the moment method, we obtain the approximate distance $\widetilde{d}_{ij}$ by solving $(d_{ij}^{o})^{1/4} \approx E[(d_{ij}^{o})^{1/4}] + \sqrt{Var((d_{ij}^{o})^{1/4})}$.}. When we do not use the RSS-based measurement model, we set $w_{ij} = 1$ for $(i,j)\in E$ and zero otherwise.

%----------------------------------------------------------
\begin{figure} [t]
 \centering    
  {\epsfig{figure=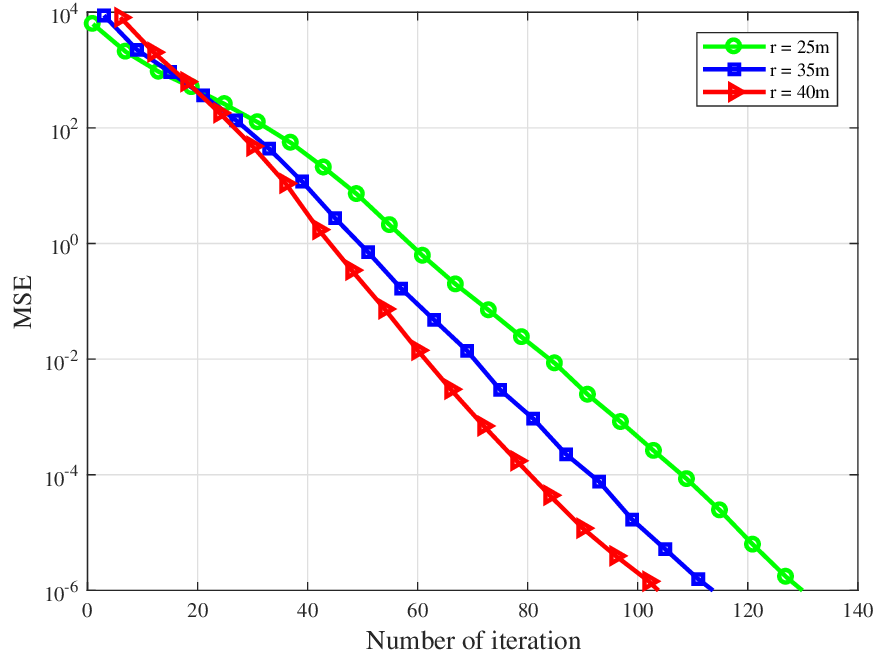, height=76mm, width=87mm}}	
  \caption {The MSE performance of LRM-CG for $k = 2$ (2-dimensional location vectors).}
\label{fig:fig3}
\end{figure}
%---------------------------------------------------------- 

%----------------------------------------------------------
\begin{figure*} [t]
 \centering
    \subfigure[]
  {\epsfig{figure=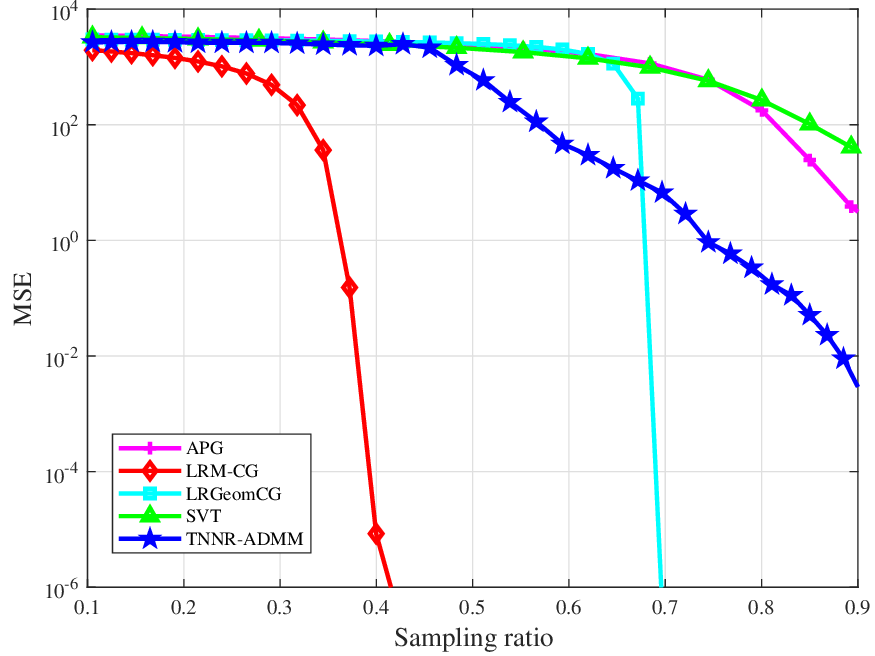, height=76mm, width=87mm}}
	\subfigure[]
  {\epsfig{figure=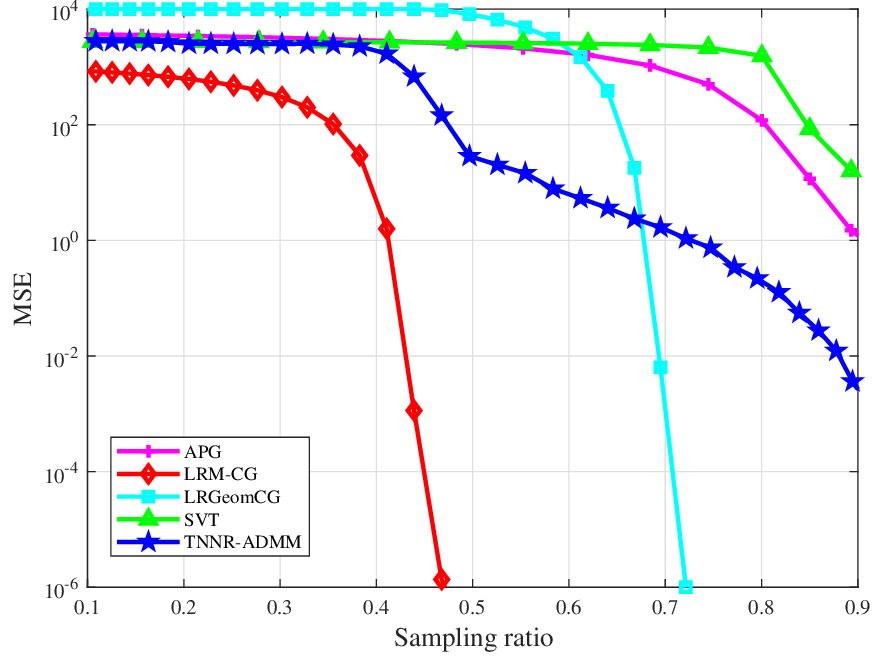, height=76mm, width=87mm}}
  \caption {The MSE performance of the matrix completion algorithms for scenario without observation error for (a) 2-dimensional and (b) 3-dimensional location vectors.}
\label{fig:fig7}
\end{figure*}
%----------------------------------------------------------

\subsection{Convergence Efficiency} 
As performance measures, we use the mean square error (MSE) and the root mean square errors (RMSE), which are defined respectively as
\begin{eqnarray}
MSE & = & \frac{1}{\sqrt{n^2-n}}\|\widehat{\mathbf{D}}-\mathbf{D}\|_F,\nonumber\\
RMSE & = & \sqrt{\frac{1}{n^2-n}\sum\limits_{i}\sum\limits_{j\neq i}(\widehat{d}_{ij}-d_{ij})^2}\nonumber .
\end{eqnarray}
Note that the number of non-trivial entries of $\mathbf{D}$ is $n^2-n$ since the diagonal elements are zero (i.e., $d_{ii} = 0$). Also, in order to compare the localization performance of the proposed algorithm, we use the mean square localization error (MSLE):
\begin{eqnarray}
\mathcal{E} & = & \frac{1}{\text{Total unknown nodes}}\sum\limits_{\text{All unknown nodes i}}\|\widehat{\mathbf{x}}_i-\mathbf{x}_i\|_2\nonumber .
\end{eqnarray} 
In Fig. \ref{fig:fig3}, we plot the log-scale MSE as a function of the number of iterations for the 2-dimensional sensor networks. Note that the results are obtained for the scenario where 200 sensor nodes are randomly distributed in $50\times 50$m$^2$ square area. We observe that the log-scale MSE decreases linearly with the number of iterations, meaning that the MSE decreases exponentially with the number of iterations. For example, if $r = 35$m, it takes about 60, 80, and 100 iterations to achieve $10^{-1}$, $10^{-3}$, and $10^{-5}$, respectively. Also, as expected, required number of iterations to achieve the given performance level decreases with the radio communication range $r$.

\subsection{Performance Evaluation}
\label{subsec:subsec3}

In this subsection, we investigate the recovery performance of LRM-CG for scenarios with and without observation error. In Fig. \ref{fig:fig7}, we plot the performance of the scenario without the observation error as a function of the sampling ratio, which is defined as the ratio of the number of observed pairwise distances to total number of pairwise distances. Here, the sampling ratio is controlled by the radio communication range $r$\footnote{In 2 and 3-dimensional Euclidean spaces, it can be shown that the sampling probability (sampling ratio) can be expressed as a non-decreasing function of $r$ (see Appendix B in Supplementary Material).}. We observe that LRM-CG outperforms conventional techniques by a large margin, achieving MSE $\leq 10^{-5}$ using 40\% of measurements. 

%----------------------------------------------------------
\begin{figure*} [t]
 \centering
    \subfigure[]
  {\epsfig{figure=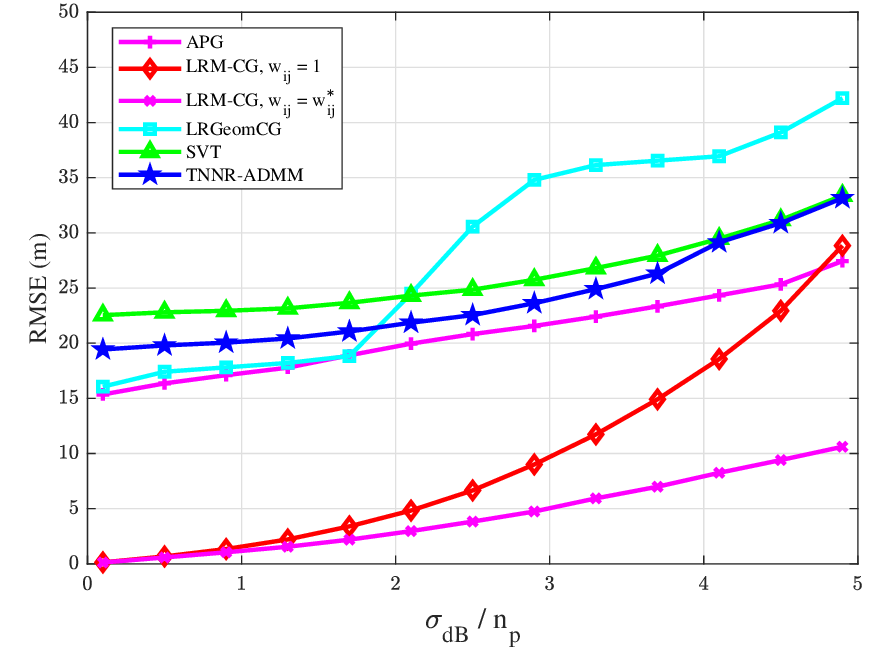, height=76mm, width=87mm}}
	\subfigure[]
  {\epsfig{figure=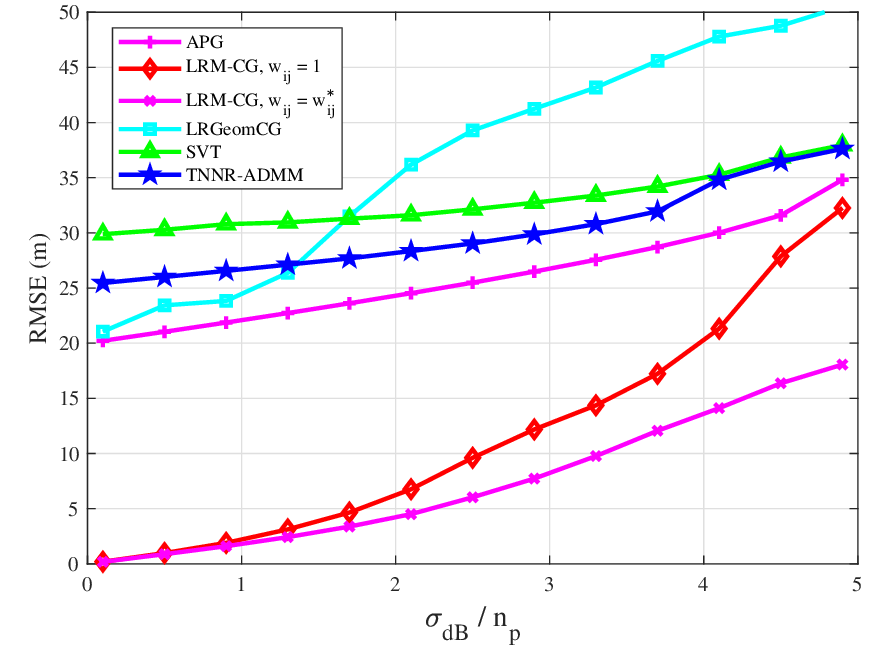, height=76mm, width=87mm}}
  \caption {The RMSE performance of the algorithms in presence of observation errors for (a) 2-dimensional and (b) 3-dimensional location vectors.}
\label{fig:fig8}
\end{figure*}
%----------------------------------------------------------

In Fig. \ref{fig:fig8}, we plot the performance of LRM-CG as a function of $\sigma_{dB}/n_p$. In this experiment, sensor nodes are randomly distributed in $50\times 50$m$^2$ square area ($k = 2$) and $50\times 50\times 50$m$^3$ cubic space ($k = 3$). We set the radio communication range $r = 30$m, resulting in 125 and 84 average connections per node for $k = 2$ and $k = 3$, respectively. While the performance of conventional matrix completion algorithms is poor (i.e., RMSE $\geq 5m$) in mid and high $\sigma_{dB}/n_p$ regime, the performance of LRM-CG is still good in small $\sigma_{dB}/n_p$ regime, achieving RMSE being less than 2.5m when $\sigma_{dB}/n_p\leq 1.5$.

\begin{figure} [t]
 \centering
  {\epsfig{figure=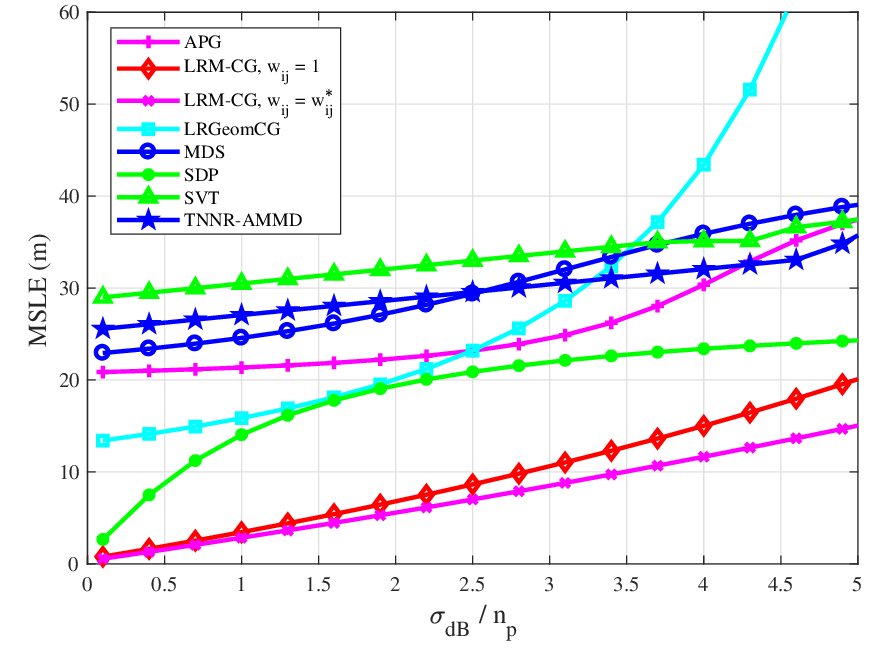, height=76mm, width=87mm}}
  \caption {The RMSLE performance of the algorithms for 3-dimensional location vectors.
   } \label{fig:fig15}
\end{figure}

We next investigate the localization performance of LRM-CG. We compare the performance of LRM-CG with the APG, LRGeomCG, SVT, TNNR-AMMD, MDS, and SDP-based algorithm \cite{localization:biswas2004}. In this experiment, 50 sensor nodes are randomly distributed in $50\times 50 \times 50$m$^3$ $(k = 3)$ and 4 anchor nodes are used to reconstruct the global node locations. The stopping threshold $\epsilon$ of LRM-CG is set to $10^{-8}$. 
Since the reconstructed matrix of the conventional matrix completion algorithm including APG, LRGeomCG, SVT, and TNNR-AMMD, is not necessarily an Euclidean distance matrix, we use the MDS technique \cite{localization:shang2003} as a post-processing to project the output matrix on the Euclidean distance matrix cone.   
In Fig. \ref{fig:fig15}, we observe that conventional localization algorithms perform poor (MSLE $\geq 5m$) for mid and high $\sigma_{dB}/n_p$ regime, but the proposed LRM-CG algorithm performs well in low $\sigma_{dB}/n_p$ regime, achieving MSLE being less than 3m for $\sigma_{dB}/n_p\leq 1$.

\begin{figure*} [t]
 \centering
    \subfigure[]
  {\epsfig{figure=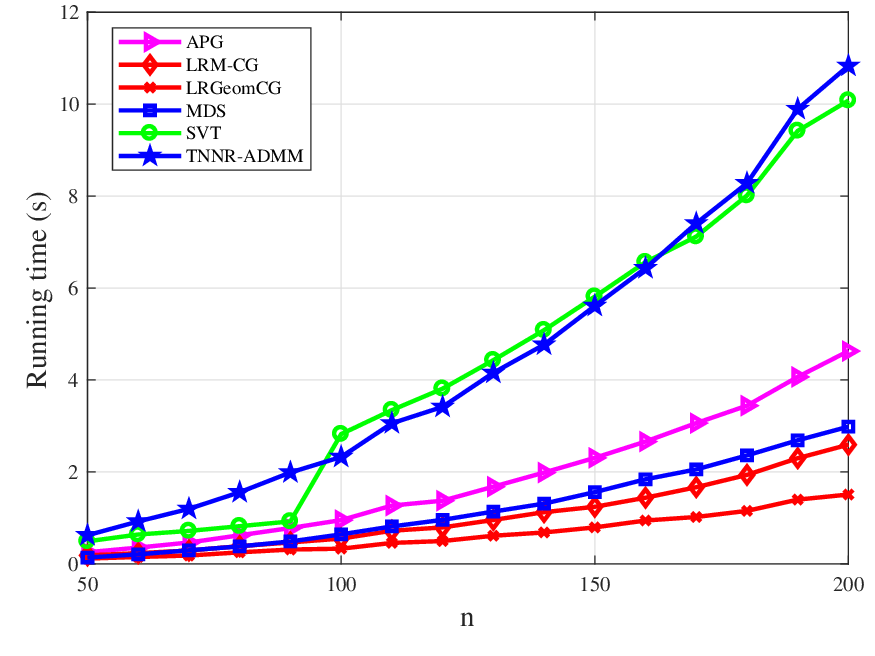, height=76mm, width=87mm}}
	\subfigure[]
  {\epsfig{figure=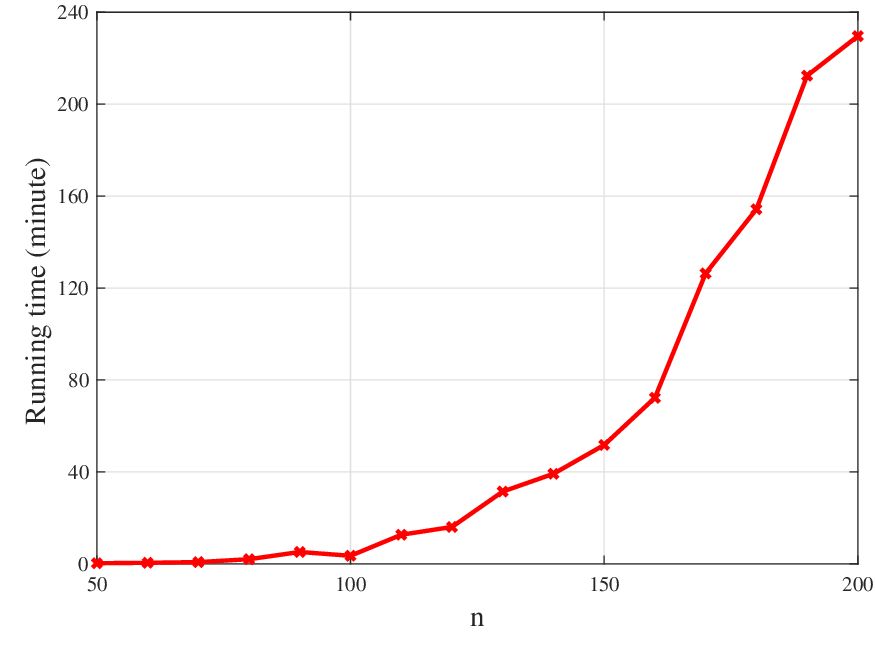, height=76mm, width=87mm}}
  \caption {Running time as a function of the number of sensor nodes: (a) the conventional matrix completion algorithms and the proposed LRM-CG and (b) SDP-based algorithm. Since the running time of SDP-based algorithm is much higher than that of the other algorithms, we separate the results into two plots.} \label{fig:fig14}
\end{figure*}

We next examine the running time complexity of the algorithms under test as a function of the number of sensor nodes. In our simulations, we set the maximum iteration number to 200 and the stopping threshold $\epsilon$ of the matrix completion algorithms to $10^{-6}$. From Fig. \ref{fig:fig14}, we observe that the running time of the SDP-based technique is fairly large since it should solve the primal and dual problems using SDPT3 solver \cite{localization:toh1999,localization:tutuncu2003}. The running time of APG, LRGeomCG, MDS, and the proposed LRM-CG is more or less similar when $n\leq 200$. In Table \ref{tab:tab005}, we summarize the computational complexity of the algorithms under test in terms of flops. We observe that the computational complexity of LRM-CG is linearly proportional to the problem size $n$ and the number of the observed distances $|E|$, and thus competitive with the conventional approaches.

\begin{table}[t]
\centering
\caption{Computational complexity of the matrix completion algorithms in recovery of $n\times n$ rank-$k$ matrix.} 
	
    \begin{tabular}{ | c | c | c | }
    \hline
    \multirow{3}{*}{ Algorithms} & \multirow{3}{*}{ \shortstack{Major \\ computation}} & \multirow{3}{*}{ \shortstack{Total computational \\ complexity \\ per iteration}} \\ 
     & &   \\  
     & &   \\ \hline 

      APG & Soft-thresholding SVD &  $\mathcal{O}(\bar{k} n^2)${\color{black}$^a$}   \\ 

      LRM-CG & Truncated EVD & $\mathcal{O}(k^2n + k|E|)$   \\
      LRGeomCG & Truncated EVD & $\mathcal{O}(k^2n + k|E|)$   \\
      MDS & Truncated EVD & $\mathcal{O}(kn^2)$   \\
      SDP & Convex operator & $\mathcal{O}(n^3)$   \\
      SVT & Soft-thresholding SVD & $\mathcal{O}(\bar{k} n^2)$  \\  
      TNNR-ADMM & Soft-thresholding SVD & $\mathcal{O}(\bar{k} n^2)$   \\  \hline    
    \end{tabular}  
    \label{tab:tab005}
    \begin{tablenotes}
        \item[a] \hspace{-2mm}{\color{black}$^a$}Note that $\bar{k}$ is the number of singular values being larger than the threshold used in the soft-thresholding based SVD technique \cite{localization:lin2009,localization:cai2010,localization:hu2013}.
    \end{tablenotes} 
\end{table}

\subsection{Outlier Problem}
\label{subsec:subsec4}

\begin{figure} [t]
 \centering
  {\epsfig{figure=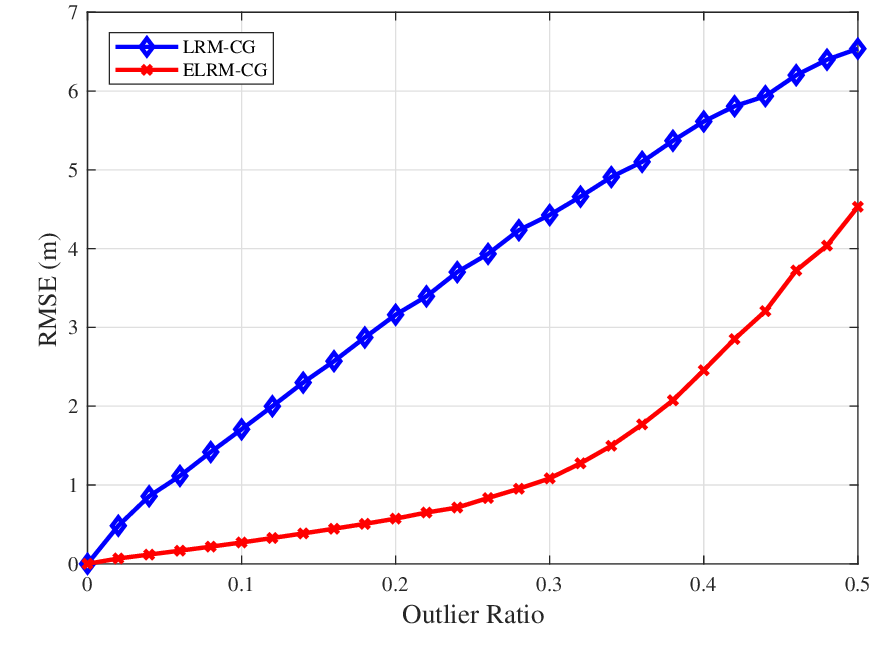, height=76mm, width=87mm}}
  \caption {The MSLE performance of LRM-CG in the presence of outliers. 
   } \label{fig:fig12}
\end{figure}

We next investigate the performance of the proposed LRM-CG algorithm and its extended version (see Section \ref{sec:sec5}) in the presence of outliers. When the outlier ratio $\theta$ is given, we randomly choose a set of the observed distances and replace this set by a set of random numbers. In this experiment, sensor nodes are randomly distributed in $50\times 50$m$^2$ square area. 
In our simulation, we consider the scenario in which the magnitude of outliers is comparable to the distance level. We could observe that the extended LRM-CG outperforms the original LRM-CG, achieving MSLE being less than 0.5m up to the 20\% outlier ratio (see Fig. \ref{fig:fig12}). 

%----------------------------------------------------------
%----------------------------------------------------------

%----------------------------------------------------------
%----------------------------------------------------------

%----------------------------------------------------------
%----------------------------------------------------------

\subsection{Real Data}
In this subsection, we examine the performance of the proposed LRM-CG algorithm using real measurements. In this simulation, we use the RSS-based measurement model in \cite{localization:costa2006}. This network consists of 44 sensor nodes randomly distributed in the $14\times 14$m$^2$ square area and the transmit signal is generated via a wideband direct-sequence spread-spectrum (DSSS) operating at a center frequency of 2.4 GHz. For a given radio communication range $r$, we assume that $d_{ij}^{o}$ is known if $d_{ij}\leq r$ and unknown otherwise.  
We observe from Table \ref{tab:tab004} that the performance of the proposed LRM-CG is comparable to the SDP techniques in \cite{localization:biswas2004, localization:guo2016}\footnote{The SDP-based techniques have various cost functions. In \cite{localization:biswas2004}, the cost function is expressed as a sum of absolute errors in terms of the observed distances while that in \cite{localization:guo2016} is a least squares function.} when $r = 9.5m$.

\begin{table*}[!t]
\centering
\caption{Localization errors with real measurements.} 
	   
    \begin{tabular}{ | c | c | c | c | c | c | }
    \hline
    \multirow{3}{*}{ $r$ (m)} & \multirow{3}{*}{ \shortstack{Average \\ connection \\ per node}} & \multicolumn{4}{c|}{MSLE (m)} \\ \cline{3-6}    
     & & \multirow{2}{*}{LRM-CG} &  \multirow{2}{*}{ELRM-CG} & \multirow{2}{*}{ \shortstack{ SDP with absolute \\ cost function \cite{localization:biswas2004} } }  &   \multirow{2}{*}{ \shortstack{ SDP with least square \\ cost function \cite{localization:guo2016} }}  \\ 
     & &  &  & & \\ \hline 
      5.5 & 14 & 5.4893 & 4.9860 & 4.5038 & 3.7241    \\  
      7.5 & 22 & 5.2796 & 4.9170 &  3.1287 & 3.3394   \\  
      9.5 & 30 & 2.9917 & 2.8620 & 2.9274 & 3.0526   \\ 
      11.5 & 37 & 2.2636 & 2.2023 & 2.6272 & 2.5151   \\ \hline    
    \end{tabular} 
    \label{tab:tab004}
\end{table*} 

\section{Conclusion}

In this paper, we have proposed an algorithm to recover the Euclidean distance matrix and the location map from partially observed distance information. In solving the Frobenius norm minimization problem with a rank constraint, we expressed the Euclidean distance matrix as a function of the low rank PSD matrix. By capitalizing on the Riemannian manifold structure for this set of matrices, we could solve the low-rank matrix completion problem using a modified nonlinear conjugate gradient algorithm. The proposed LRM-CG algorithm preserves the low rank structure of this reconstructed matrix. We have shown from the recovery condition analysis that the proposed LRM-CG algorithm converges to the original Euclidean distance matrix in the sampling space under the extended Wolfe's conditions. We have also demonstrated from the numerical experiments that LRM-CG outperforms the conventional matrix completion techniques by a large margin, achieving MSE $\leq 10^{-5}$ using 40\% of measurements. We also proposed an extended version of LRM-CG to control the outliers and demonstrated the effectiveness of the proposed scheme in the realistic environments with outliers. Given the importance of the location-aware applications and services in the IoT era, we believe that the proposed LRM-CG algorithm will be a useful tool for various localization scenarios. While our work focused primarily on the network localization scenario, extension to the distributed network scenarios would also be interesting direction worth pursuing.



%\section*{Acknowledgment}

\begin{appendices}

\section{Proof of Theorem \ref{lm:lm006}} 
\label{app:appB}
\begin{proof}
Since $\nabla_{\mathbf{Y}}f(\mathbf{Y})$ is interpreted as a matrix whose inner product with an arbitrary matrix $\mathbf{H}$ becomes the Frechet differential $\text{D}f(\mathbf{Y})[\mathbf{H}]$ of $f$ at $\mathbf{Y}$, it is convenient to compute $\nabla_{\mathbf{Y}}f(\mathbf{Y})$ as a unique element of $\mathbb{R}^{n\times n}$ that satisfies 
\begin{equation}
<\nabla_{\mathbf{Y}}f(\mathbf{Y}),\mathbf{H}> = \text{D}f(\mathbf{Y})[\mathbf{H}],
\label{eq:eq527}
\end{equation}
for all $\mathbf{H}$. We first compute $\text{D}f(\mathbf{Y})[\mathbf{H}]$ and then use (\ref{eq:eq527}) to obtain the expression of $\nabla_{\mathbf{Y}}f(\mathbf{Y})$. Let $h(\mathbf{R}) = \frac{1}{2}\|\mathbf{R}\|_F^2$ and $k(\mathbf{Y}) = \mathbf{W}\odot(\mathcal{P}_{E}\circ g)(\mathbf{Y})-\mathbf{W}\odot\mathcal{P}_E(\mathbf{D}_{o})$, then $f(\mathbf{Y}) = h(k(\mathbf{Y})) = (h\circ k)(\mathbf{Y})$. Using matrix calculus, it is not hard to show that 
\begin{eqnarray}
\text{D}f(\mathbf{Y})[\mathbf{H}] 
& = & <2\text{eye}(\text{Sym}(\mathbf{W}\odot k(\mathbf{Y}))\mathbf{1})\nonumber\\
&  &-2\text{Sym}(\mathbf{W}\odot k(\mathbf{Y})),\mathbf{H}>
\label{eq:eq734}
\end{eqnarray} 
From (\ref{eq:eq527}) and (\ref{eq:eq734}), we have $\nabla_\mathbf{Y}f(\mathbf{Y}) = 2\text{eye}(\text{Sym}(\mathbf{W}\odot k(\mathbf{Y}))\mathbf{1})-2\text{Sym}(\mathbf{W}\odot k(\mathbf{Y}))$, which is the desired result.

\end{proof}

%=====================================

\section{Proof of Lemma \ref{lm:lm308}} 
\label{app:appAJ}
\begin{proof}
First, a lower bound of $\|\nabla_\mathbf{Y}f(\mathbf{Y}_i)\|_F$ is given by
\begin{eqnarray}
\|\nabla_\mathbf{Y}f(\mathbf{Y}_i)\|_F^2 & \substack{(a)\\ =} & \|2\text{eye}(\mathbf{R}_i\mathbf{1})-2\mathbf{R}_i\|_F^2\nonumber\\
& \substack{(b)\\ =} & \|2\text{eye}(\mathbf{R}_i\mathbf{1})\|_F^2+\|2\mathbf{R}_i\|_F^2\nonumber\\
& \geq & \|2\mathbf{R}_i\|_F^2,
%& = & 4\|\mathcal{P}_E(\mathbf{D}_i)-\mathcal{P}_E(\mathbf{D})\|_F^2,
\label{eq:eq713}
\end{eqnarray} 
where (a) is from (\ref{eq:eq562}) and (b) is from the fact that diagonal entries of $\mathbf{R}_j$ are all zeros and $\text{eye}(\mathbf{R}_j\mathbf{1})$ is a diagonal matrix. That is, positions of nonzero elements in $\text{eye}(\mathbf{R}_i\mathbf{1})$ and $\mathbf{R}_i$ are disjoint. An upper bound is obtained as follows.
\begin{eqnarray}
\|\nabla_\mathbf{Y}f(\mathbf{Y}_i)\|_F & \leq & \|2\text{eye}(\mathbf{R}_i\mathbf{1})\|_F+\|2\mathbf{R}_i\|_F\nonumber\\
& \substack{(a)\\ \leq} & \|2\mathbf{R}_i\mathbf{1}\|_2+\|2\mathbf{R}_i\|_F \nonumber \\ 
& \substack{(b)\\ \leq} & 2\|\mathbf{R}_i\|_F\|\mathbf{1}\|_2+2\|\mathbf{R}_i\|_F \nonumber \\ 
& \leq & (2\sqrt{n}+2)\|\mathbf{R}_i\|_F,
%& \leq & (2\sqrt{n}+2)\|\mathcal{P}_E(\mathbf{D}_i)-\mathcal{P}_E(\mathbf{D})\|_F,
\label{eq:eq715}
\end{eqnarray}
where (a) is because $\|\text{eye}(\mathbf{b})\|_F = \|\mathbf{b}\|_2$ for any vector $\mathbf{b}$, and (b) is because $\|\mathbf{A}\mathbf{b}\|_2\leq \|\mathbf{A}\|_F\|\mathbf{b}\|_2$ for any matrix $\mathbf{A}$ and any vector $\mathbf{b}$. By combining (\ref{eq:eq713}) and (\ref{eq:eq715}), and noting that $\|\mathbf{R}_i\|_F = \|\mathcal{P}_E(\mathbf{D}_i)-\mathcal{P}_E(\mathbf{D})\|_F$, we obtain the desired result.  
\end{proof}

%=====================================
\section{Proof of Lemma \ref{lm:lm301}} 
\label{app:appAA}

\begin{proof}
From \textbf{A3}, we have $\|\text{grad}f(\mathbf{Y}_i)\|_F^2  \geq  \frac{1}{c^2}\|\nabla_\mathbf{Y}f(\mathbf{Y}_i)\|_F^2$.
Now, what remains is to show that $\|\nabla_\mathbf{Y}f(\mathbf{Y}_i)\|_F^2\geq 8f(\mathbf{Y}_i)$. Indeed, from Lemma \ref{lm:lm006}, we have
%\begin{equation}
%\|\nabla_\mathbf{Y}f(\mathbf{Y}_i)\|_F^2 = \|\text{eye}((\mathbf{R}+\mathbf{R}^T)\mathbf{1}) - 2\mathbf{R}\|_F^2, \nonumber
%\end{equation}  
%where $\mathbf{R} = \mathcal{P}_E(g(\mathbf{Y}_i)) - \mathcal{P}_E(\mathbf{D})$. Noting that $\mathbf{R}$ is symmetric with zero diagonal entries $r_{ii} = 0$, we have
\begin{align}
\frac{1}{4}\|\nabla_\mathbf{Y}f(\mathbf{Y}_i)\|_F^2 & =  \frac{1}{4}\|\text{eye}((\mathbf{R}+\mathbf{R}^T)\mathbf{1}) - 2\mathbf{R}\|_F^2\notag\\
& =  \|\text{eye}(\mathbf{R}\mathbf{1})\|_F^2 + \|\mathbf{R}\|_F^2\notag\\
& - 2<\text{eye}(\mathbf{R}\mathbf{1}),\mathbf{R}>\notag\\
& =  \|\mathbf{R}\mathbf{1}\|_2^2 + \|\mathbf{R}\|_F^2 -2\sum\limits_i\left(\sum\limits_{j}r_{ij}\right)r_{ii}\notag\\
& =  \|\mathbf{R}\mathbf{1}\|_2^2 + \|\mathbf{R}\|_F^2 \notag\\
& \geq  \|\mathbf{R}\|_F^2, 
\label{eq:eq303}
\end{align}
where $\mathbf{R} = \mathcal{P}_E(g(\mathbf{Y}_i)) - \mathcal{P}_E(\mathbf{D})$ is symmetric with zero diagonal entries $r_{ii} = 0$.
Noting that $\|\mathbf{R}\|_F^2 = 2f(\mathbf{Y}_i)$, we obtain the desired result.
\end{proof} 

\end{appendices}



\bibliographystyle{IEEEtran}

%\bibliography{IEEEabrv,localization}

\begin{thebibliography}{10}
\bibitem{localization:luongITA}
L.~Nguyen, S.~Kim, and B.~Shim, ``Localization in internet of things network:
  Matrix completion approach,'' in \emph{Proc. Inform. Theory Applicat. Workshop},
  2016.

\bibitem{localization:luongICCC}
L.~Nguyen and B.~Shim, ``Localization of internet of things network via
  euclidean distance matrix completion,'' in \emph{Proc. IEEE/CIC Int. Conf.
  Commun. China (ICCC).}, 2016.

\bibitem{localization:delamo2015}
M.~Delamo, S.~Felici-Castell, J.~J. Perez-Solano, and A.~Foster, ``Designing an
  open source maintenance-free environmental monitoring application for
  wireless sensor networks,'' \emph{J. Syst. Softw.}, vol. 103, pp. 238--247,
  May 2015.

\bibitem{localization:sunwoo2014}
S.~Lee, B.~Koo, and S.~Kim, ``{RAPS}: reliable anchor pair selection for
  range-free localization in anisotropic networks,'' \emph{IEEE Commun. Lett.},
  vol.~18, pp. 1403--1406, 2014.

\bibitem{localization:hackmann2014}
G.~Hackmann, W.~Guo, G.~Yan, Z.~Sun, C.~Lu, and S.~Dyke, ``Cyber-physical
  codesign of distributed structural health monitoring with wireless sensor
  networks,'' \emph{IEEE Trans. Parallel Distrib. Syst.}, vol.~25, pp. 63--72,
  Jan. 2014.

\bibitem{localization:pal2010}
A.~Pal, ``Localization algorithms in wireless sensor networks: Current
  approaches and future challenges,'' \emph{Netw. Protocols Algorithms},
  vol.~2, no.~1, pp. 45--74, 2010.

\bibitem{localization:hodge2014}
V.~J. Hodge, S.~O'Keefe, M.~Weeks, and A.~Moulds, ``Wireless sensor networks
  for condition monitoring in the railway industry: A survey,'' \emph{IEEE
  Trans. Intell. Transp. Syst.}, vol.~16, pp. 1088--1106, Jun. 2015.

\bibitem{localization:rawat2014}
P.~Rawat, K.~D. Singh, H.~Chaouchi, and J.~M. Bonnin, ``Wireless sensor
  networks: a survey on recent developments and potential synergies,'' \emph{J.
  Supercomput.}, vol.~68, no.~1, pp. 1--48, Apr. 2014.

\bibitem{localization:aspnes2006}
J.~Aspnes, T.~Eren, D.~Goldenberg, A.~Morse, W.~Whiteley, Y.~Yang, B.~Anderson,
  and P.~Belhumeur, ``A theory of network localization,'' \emph{IEEE Trans.
  Mobile Comput.}, vol.~5, no.~12, pp. 1663--1678, Dec. 2006.

\bibitem{localization:shang2003}
Y.~Shang, W.~Ruml, Y.~Zhang, and M.~Fromherz, ``Localization from mere
  connectivity,'' in \emph{Proc. ACM Symp. Mobile Ad Hoc Netw. Comput.},
  Annapolis, Jun. 2003, pp. 201--212.

\bibitem{localization:parker2007}
R.~Parker and S.~Valaee, ``Vehicular node localization using
  received-signal-strength indicator,'' \emph{IEEE Trans. Veh. Technol.},
  vol.~56, pp. 3371--3380, Nov. 2007.

\bibitem{localization:dardari2008}
D.~Dardari, C.-C. Chong, and M.~Z. Win, ``Threshold-based time-of-arrival
  estimators in uwb dense multipath channels,'' \emph{IEEE Trans. Commun.},
  vol.~56, pp. 1366--1378, Aug. 2008.

\bibitem{localization:zhang2005}
Y.~Zhang and J.~Zha, ``Indoor localization using time difference of arrival and
  time-hopping impulse radio,'' \emph{IEEE Int. Symp. Commun. Inform.
  Technol.}, pp. 964--967, Oct. 2005.

\bibitem{localization:guo2016}
X.~Guo, L.~Chu, and X.~Sun, ``Accurate localization of multiple sources using
  semidefinite programming based on in complete range matrix,'' \emph{IEEE
  Sensors J.}, vol.~16, no.~13, pp. 5319--5324, July 2016.

\bibitem{localization:biswas2004}
P.~Biswas and Y.~Ye, ``Semidefinite programming for ad hoc wireless sensor
  network localization,'' in \emph{Proc. 3rd Int. Symp. Inform. Process. Sensor
  Networks. ACM.}, pp. 46--54, 2004.

\bibitem{localization:candes2009}
E.~J. Candes and B.~Recht, ``Exact matrix completion via convex optimization,''
  \emph{Found. Comput. Math.}, vol.~6, pp. 717--772, 2009.

\bibitem{localization:cai2010}
J.-F. Cai, E.~J. Candes, and Z.~Shen, ``A singular value thresholding algorithm
  for matrix completion,'' \emph{SIAM J. Optimiz.}, vol.~20, no.~4, pp.
  1956--1982, 2010.

\bibitem{localization:lin2010}
Z.~Lin, M.~Chen, and Y.~Ma., ``The augmented lagrange multiplier method for
  exact recovery of corrupted low-rank matrices,'' \emph{arXiv:1009.5055},
  2010.

\bibitem{localization:mishra2011}
B. Mishra, G.~Meyer, and R.~Sepulchre, 
\newblock ``Low-rank optimization for distance matrix completion,'' 
\newblock in \emph{Proc. 50th IEEE Conf. Decision Control Eur. Control Conf. (CDC-ECC)}, pp. 4455--4460, Dec. 2011.

\bibitem{localization:jain2013}
P. Jain, P.~Netrapalli, and S.~Sanghavi, 
\newblock ``Low-rank matrix completion using alternating minimization,'' 
\newblock in \emph{Proc. 45th Annu. ACM Symp. Theory Comput.}, pp. 665–674, 2013.

\bibitem{localization:dattorro2005}
J.~Dattorro, \emph{Convex Optimization and Euclidean Distance Geometry}, USA: Meboo Publishing, 2005.

\bibitem{localization:lin2009}
Z.~Lin, A.~Ganesh, J.~Wright, L.~Wu, M.~Chen, and Y.~Ma, ``Fast convex
  optimization algorithms for exact recovery of a corrupted low-rank matrix,'' in
  \emph{Proc. Int. Workshop Comput. Adv. Multi-Sensor Adapt. Process.}, pp.
  1--18, 2009.

\bibitem{localization:absil2008}
P.~A. Absil, R.~Mahony, and R.~Sepulchre, \emph{Optimization Algorithms on
  Matrix Manifolds}, Princeton Univ. Press, 2008.

\bibitem{localization:jianwu2009}
Z.~Jianwu and Z.~Lu, ``Research on distance measurement based on rssi of
  zigbee,'' \emph{ISECS Int. Colloq. Computing, Commun., Control, and Manage.},
  pp. 210--212, 2009.

\bibitem{localization:helmke1994}
U.~Helmke and J.~B. Moore, \emph{Optimization and Dynamical Systems}, London: Springer-Verlag, 1994.

\bibitem{localization:bart2013}
B.~Vandereycken, ``Low-rank matrix completion by riemannian optimization,''
  \emph{SIAM J. Optimiz.}, vol.~23, no.~2, pp. 1214--1236, 2013.

\bibitem{localization:absil2012}
P.~A. Absil and J.~Malick, ``Projection-like retractions on matrix manifolds,''
  \emph{SIAM J. Optimiz.}, vol.~22, pp. 135--158, 2012.

\bibitem{localization:dai2011}
Y.~H. Dai, ``Nonlinear conjugate gradient methods,'' \emph{Wiley Encyclopedia
  of Operations Research and Manage. Sci.}, 2011.

\bibitem{localization:hager2005}
W.~W. Hager and H.~Zhang, ``A new conjugate gradient method with guaranteed
  descent and an efficient line search,'' \emph{SIAM J. Optimiz.}, no.~1, pp.
  170--192, 2005.

\bibitem{localization:wolfe1969}
P.~Wolfe, ``Convergence conditions for ascent methods,'' \emph{SIAM Rev.},
  no.~2, pp. 226--235, 1969.

\bibitem{localization:sato2015}
H.~Sato and T.~Iwai, ``A new, globally convergent riemannian conjugate gradient
  method,'' \emph{Optim. J. Math. Program. Oper. Res.}, no.~4, pp. 1011--1031,
  2015.

\bibitem{localization:yang2013}
Z.~Yang, C.~Wu, T.~Chen, Y.~Zhao, W.~Gong, and Y.~Liu, ``Detecting outlier
  measurements based on graph rigidity for wireless sensor network
  localization,'' \emph{IEEE Trans. Veh. Technol.}, vol.~62, no.~1, pp.
  374--383, Jan. 2013.

\bibitem{localization:hu2013}
Y.~Hu, D.~Zhang, J.~Ye, X.~Li, and X.~He, ``Fast and accurate matrix completion
  via truncated nuclear norm regularization,'' \emph{IEEE Trans. Pattern Anal.
  Mach. Intell.}, no.~9, pp. 2117--2130, Sep. 2013.

\bibitem{localization:costa2006}
J.~A. Costa, N.~Patwari, and I.~A.~O.~Hero, ``Distributed
  weighted-multidimensional scaling for node localization in sensor networks,''
  \emph{ACM Trans. Sensor Netw.}, vol.~2, no.~1, pp. 39--64, 2006.

\bibitem{localization:toh1999}
K.~C. Toh, M.~J. Todd, and R.~H. Tutuncu, ``Sdpt3---a matlab software package
  for semidefinite programming,'' \emph{Optimization Methods and Software},
  vol.~11, pp. 545--581, 1999.

\bibitem{localization:tutuncu2003}
R.~H. Tutuncu, K.~C. Toh, and M.~J. Todd, ``Solving semidefinite quadratic
  linear programs using sdpt3,'' \emph{Math. Programming Ser. B}, vol.~95, pp.
  189--217, 2003.

\end{thebibliography}
%\begin{thebibliography}{1}
%
%
%\end{thebibliography}

\end{document}